\documentclass{aa}  

\usepackage{graphicx}
\usepackage{txfonts}
\usepackage{natbib}
\def\msun{$M_{\odot}$}
\newcommand{\angstrom}{\mbox{\normalfont\AA}}

\begin{document}

   \title{
   The environment of supernova remnant VRO 42.05.01 as probed with IRAM 30m molecular line observations}
   
      \author{M.~Arias \inst{1} \and
      V.~Dom{\v{c}}ek \inst{1,2} \and
            P.~Zhou \inst{1} \and
    J.~Vink \inst{1,2,3} 
          }

   \institute{
   Anton Pannekoek Institute for Astronomy, University of Amsterdam, Science Park 904, 1098 XH Amsterdam, The Netherlands \and
   GRAPPA, University of Amsterdam, Science Park 904, 1098 XH Amsterdam, The Netherlands \and 
   SRON, Netherlands Institute for Space Research, Utrech, The Netherlands
             }



   \date{}

 
  \abstract
   {The environment of supernova remnants (SNRs) is a key factor in their evolution, particularly at later stages of their existence.
    Mixed-morphology (MM) SNRs have a peculiar centre-filled X-ray shape that remains enigmatic.
    It is often assumed that they evolve in very dense environments, and that the X-ray morphology is due
    to interactions between the SNRs and their surroundings.}
   {We aim to determine whether VRO 42.05.01 is embedded in, and interacting with, a dense molecular environment. 
   We also aim to understand the multi-wavelength emission from the environment of this SNR, and whether the interstellar material can be
   responsible for the the MM nature of the source, and for its strange radio and optical shape.}
   {We used the IRAM telescope in Pico Veleta, Spain, to search for signs of interaction between
   the SNR and neighbouring molecular clouds. We observed a region of $26'\times14'$ towards the west of VRO 42.05.01 and a region 
   of $8'\times4'$  towards the north of the remnant
   in the $^{12}$CO $J=1-0$, $^{13}$CO $J=1-0$, and $^{12}$CO $J=2-1$ 
   transitions with the EMIR receiver. We made maps of the properties of the observed molecular clouds (peak temperatures, central velocities, velocity dispersions), as well
   as maps of column density along the line of sight, and ratio of the $^{12}$CO $J=2-1$ to $^{12}$CO $J=1-0$ transitions. We also analyse archival optical, infrared, and radio spectroscopic
   data for other hints on the nature of the medium.}
   {We do not find conclusive physical proof that the SNR is interacting with the few, clumpy molecular clouds that surround it in the region of our
   observations, although there is some suggestion of such interaction (in a region outside our map) from infrared emission. 
   We find that there is a velocity gradient in one of the molecular clouds that is consistent with a stellar wind blown by a $12-14$~\msun\
   progenitor star. We reassess the literature distance to VRO 42.05.01, and propose
   that it has a local standard of rest velocity of $-6$ km\,s$^{-1}$, and that it is located  $1.0\pm0.4$~kpc away (the earlier distance value was $4.5\pm1.5$~kpc).
   We find that a dust sheet intersects VRO 42.05.01 and is possibly related to its double shell-shaped morphology.
   }
   {}


   \maketitle
%

\section{Introduction}

Core-collapse supernova explosions often occur near massive molecular clouds (MCs), the sites of stellar birth.
Moreover, massive stars affect their environment through radiation and
winds. The complex environment in which a star dies can have a profound effect on the evolution of its supernova remnant (SNR).
The effect is of particular importance at late times, when the remnant has swept up many times over the mass of its progenitor,
and the original imprint of the explosion is lost.

There are several ways to establish interaction between a SNR and neighbouring MCs \citep{jiang10}; a common
one is through molecular observations of CO at different transitions and isotopes \cite[e.g.][]{huang86}. In this
work we present new molecular observations in the direction of SNR VRO 42.05.01 (G 166.0 +4.3), a peculiarly shaped mixed-morphology (MM)
SNR in the direction of the Galactic anti-centre.

Mixed-morphology supernova remnants \citep{rho98} are a class of supernova remnants that display a shell-shape
in the radio, but are centre-filled in the X-rays. The X-ray emission is thermal in nature and not caused by a pulsar wind. 
These sources tend to be associated with the denser parts of the interstellar medium (ISM), and are more likely
to be interacting with nearby MCs than SNRs are in general \cite[see Table 4 in][]{vink12}.
There is no consensus in the field over what the mechanism is by which SNRs become MM SNRs, but the two
main models \citep{white91,cox99} invoke a dense environment in which the SNR evolves. According to the former, there are
dense clouds nearby the explosion site that survive the shock passage, are evaporating inside the hot interior
of the SNR, and are responsible for the X-ray emission. The latter attributes the difference in radio and X-ray morphology 
to the interaction of the shock with a dense ISM: as the shock becomes radiative, the shell will no longer emit X-rays.
In addition to these there are other models,
such as the reflected shock model \citep{chen08}, which attributes the MM to a dense wall that reflects the forward shock
back into the direction of the ejecta, or the conduction phase model \citep{kawasaki05}, which suggests that the MM
is a stage in the life of all SNRs.

\begin{figure}
\includegraphics[width=0.95\columnwidth]{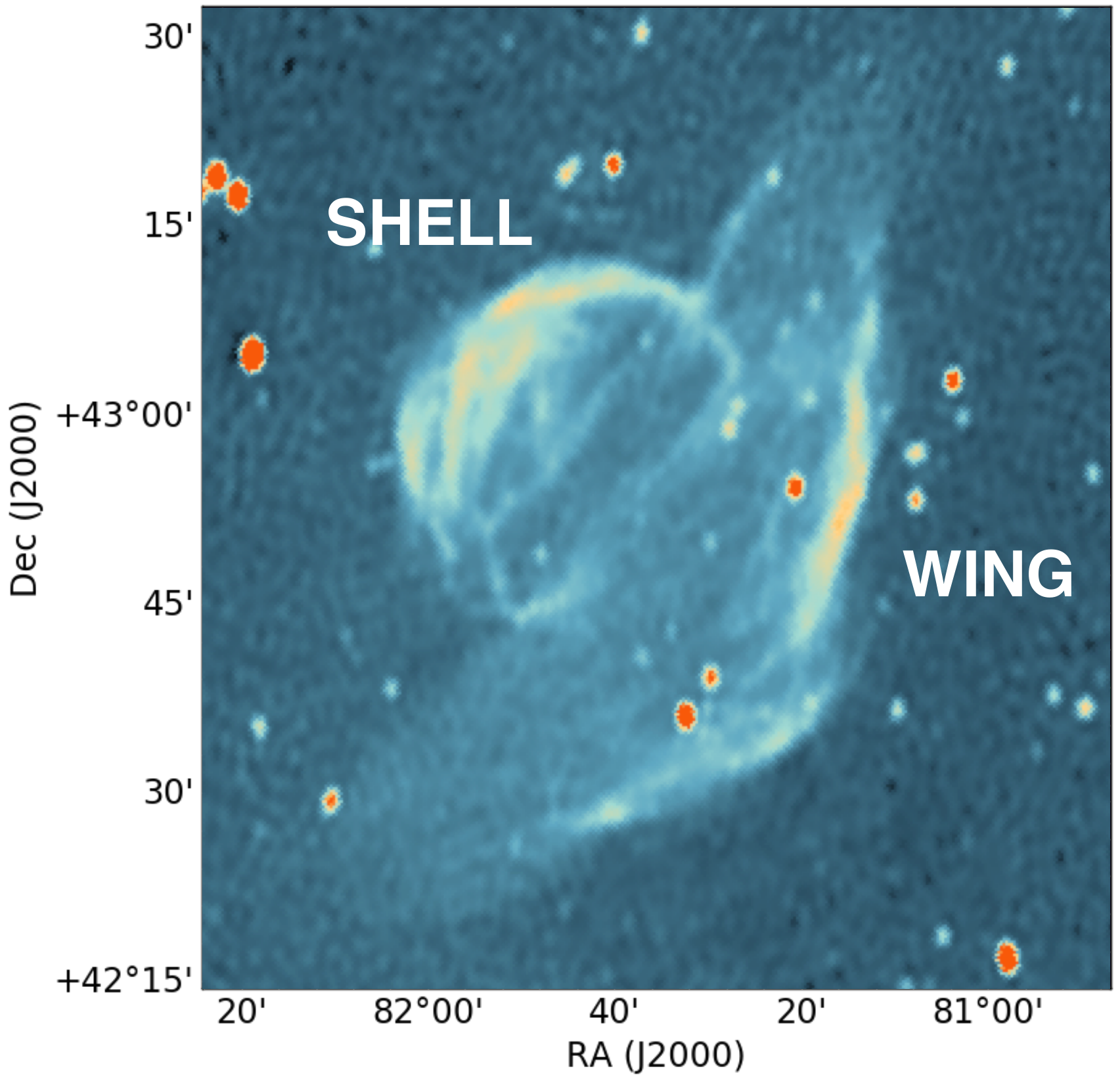}
      \caption{VRO 42.05.01 at 1420 MHz. This map taken from the CGPS. Labelled are the regions referred to throughout the text
      as \lq shell' (the small semicircle) and \lq wing' (the large triangle). We note that a sharp boundary separates the two regions.}
         \label{vro1420}
\end{figure}

VRO 42.05.01 (Fig. \ref{vro1420}) is a relatively well-studied SNR with a puzzling shape. It is formed by two structures that intersect in a sharp
line: the semicircular, smaller \lq shell', and the triangular \lq wing' (see Fig. \ref{vro1420}). It has been studied in the X-rays \citep{burrows94, guo97,bocchino09,matsumura17},
in the optical \citep{vandenbergh73,fesen85,boumis16}, in the radio continuum \citep{landecker82,pineault85,leahy05,arias19}, and in H I \citep{landecker89}.
It has also been detected in GeV $\gamma$-rays \citep{araya13}, although this detection could be due to a chance alignment with an extragalactic source.

For a source with as unusual a morphology as VRO 42.05.01, it is natural to think that the environment must
have had a significant role in its evolution. The most comprehensive model of the source to date is that of \cite{pineault87},
who proposed that the supernova occurred in a warm medium of intermediate density (which surrounds the shell)
and that a part of the remnant (the wing) broke out into a diffuse interstellar cavity, possibly blown by a stellar wind or
one or more earlier SNRs. In \cite{arias19} we elaborated on this model, suggesting that the sharp triangular cavity
of the wing could be due to the progenitor star exploding as it was moving supersonically into the more rarefied
medium, forming a bow shock whose structure remains visible in the SNR.

Our aim in this paper is to understand the environment of VRO 42.05.01 and how it has affected the
remnant's morphology. How a particular remnant developed its shape is anecdotal; 
we are more generally trying to understand the process by which this remnant became of mixed-morphology,
and the role of its medium in confining the X-ray emitting material in the interior of the radio shell.

\section{Data}

\begin{figure*}
\includegraphics[width=\textwidth]{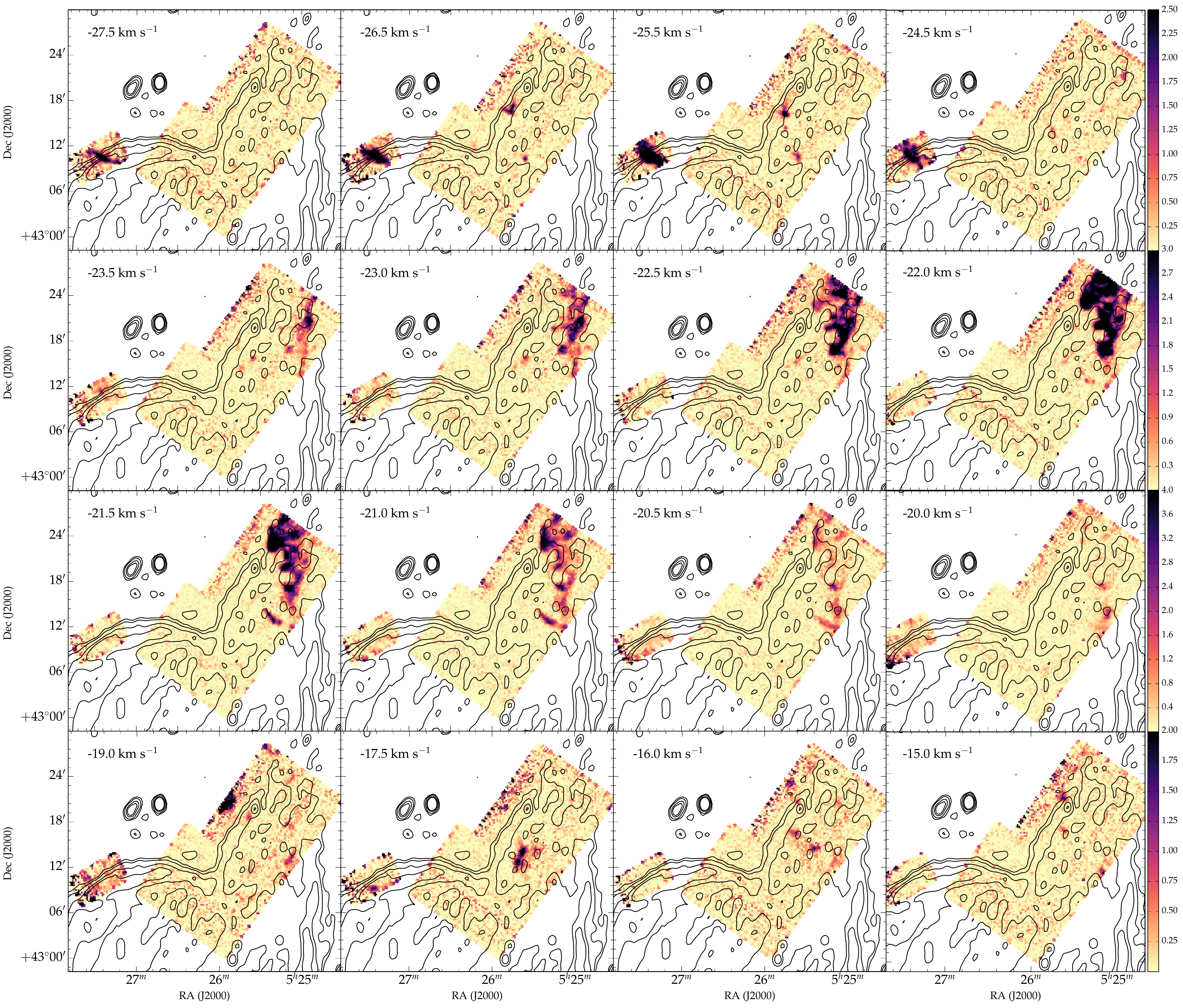}
      \caption{$T_{\mathrm{mb}}$ (K) for the $^{12}$CO $J=1-0$ transition for the distant molecular cloud components, corresponding to velocities between 
      $-27.5 \, \mathrm{~km\,s}^{-1}$ and $-15.0 \, \mathrm{~km\,s}^{-1}$. Overlaid are the radio contours at 1420 MHz
      in units of brightness temperature at 4.8, 5.0, 5.5, and 6.0 K.
      The velocities are indicated in the upper left corner of each panel. All figures in the same row share the same colour bar, plotted at the
      left end of each row.}
         \label{molec_data_dist}
\end{figure*}

\begin{figure*}{
         \centering
            \includegraphics[width=0.9\textwidth]{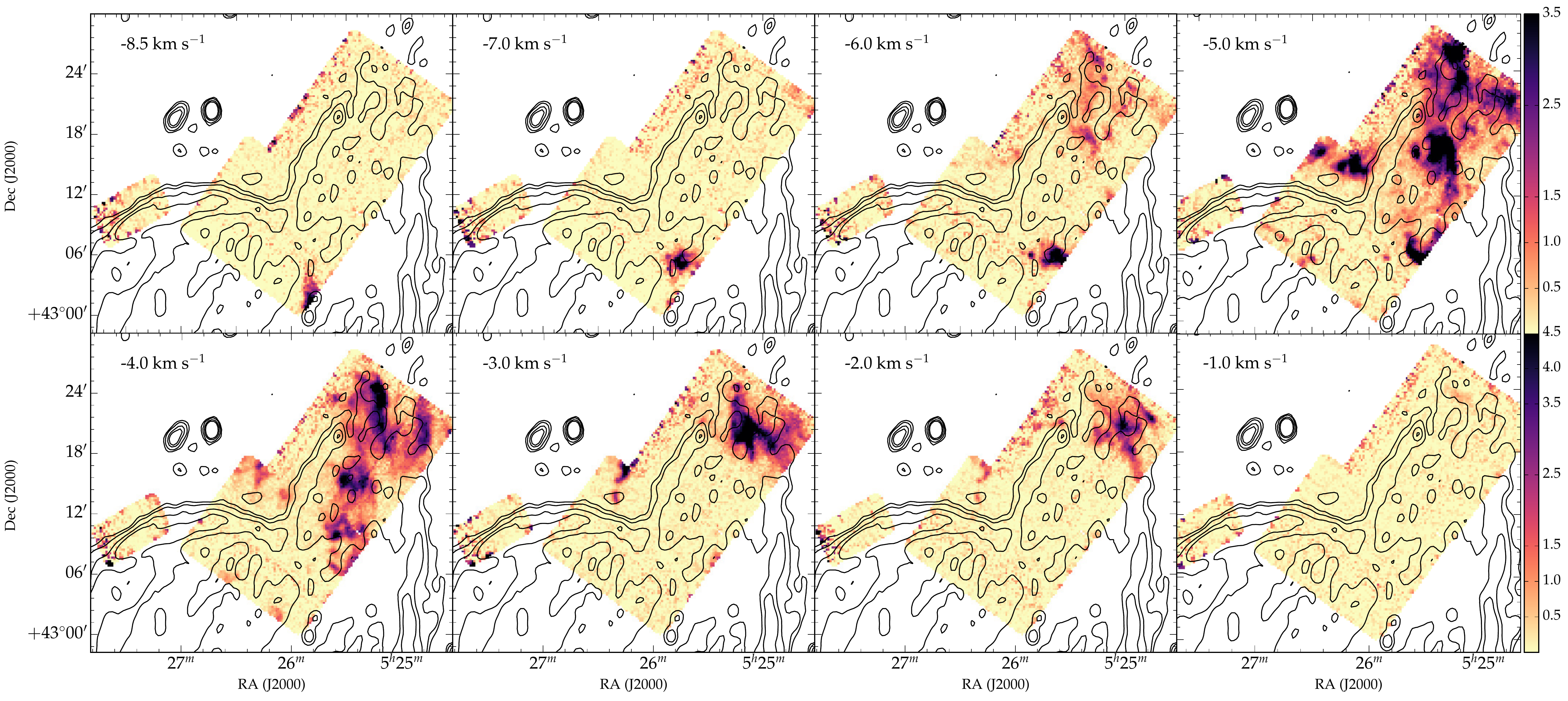}
       \caption{$T_{\mathrm{mb}}$ (K) for the $^{12}$CO $J=1-0$ transition
       for the close molecular cloud components, corresponding to velocities between 
      $-8.5 \, \mathrm{km\,s}^{-1}$ and $-1.0 \, \mathrm{km\,s}^{-1}$. Overlaid are the radio contours at 1420 MHz
      in units of brightness temperature at 4.8, 5.0, 5.5, and 6.0 K.
      The velocities are indicated in the upper left corner of each panel. All figures in the same row share the same colour bar, plotted at the
      left end of each row.}
         \label{molec_data_local}}
   \end{figure*}

\subsection{IRAM 30 m observations}

We performed 3 and 1 mm heterodyne observations toward the western side of VRO 42.05.01 with the IRAM 30 m telescope,
over several days in the summer and fall of 2018 (3, 4, and 5 July, and 24, 25, and 26 October)
under project 045-18 (PI: Mar\'ia Arias).
The observations amounted to approximately 23 hours on-source,
with varying weather and elevation conditions, resulting in uneven noise levels in our map.

We simultaneously mapped $^{12}$CO $J=1-0$ (at 115.271 GHz), $^{13}$CO $J=1-0$ (at 110.201 GHz), and $^{12}$CO $J=2-1$ 
(at 230.540 GHz) emission toward VRO 42.05.01 and its vicinity, in the
on-the-fly mode with the Eight MIxer Receiver (EMIR). 
The fast Fourier transform spectrometers (FTSs) were tuned to the 200 kHz resolution, which provided a velocity resolution of 
0.25 km s$^{-1}$ for the $^{12}$CO $J = 2-1$ line, and 0.5 km s$^{-1}$ for the $^{12}$CO $J = 1-0$  line. 
The beam efficiencies were 0.78 and 0.58 for the $J=1-0$, and $J=2-1$ transitions, respectively.

The data were then converted to $^{12}$CO $J=1-0$, $^{12}$CO $J=2-1$, and $^{13}$CO $J=1-0$ main beam temperature 
($T_\mathrm{mb}$) cubes with a grid spacing of 11.75\arcsec\ and a velocity resolution of 0.5 km s$^{-1}$. 
The final map has 300 channels, spanning the velocity range from $-99.5$~km~s$^{-1}$ to  $+50$~km~s$^{-1}$. The data
were reduced with the GILDAS/CLASS package\footnote{\url{http://www.iram.fr/IRAMFR/GILDAS/}}.

The molecular data in $T_{\mathrm{mb}}$ (K) for the $^{12}$CO $J=1-0$ transition are plotted in Figs. \ref{molec_data_dist} and \ref{molec_data_local}. 
The molecular data for the $^{12}$CO $J=2-1$ and $^{13}$CO $J=1-0$ transitions
are plotted in the appendix (Figs. \ref{12CO21}, \ref{13CO10}, and \ref{local}). Finally, the rms noise for each transition,
for each pixel, is also in the appendix (Fig. \ref{rms_noise}).

\subsection{Other archival data used for this work}

In this paper we tried to put our IRAM molecular observations in a multi-wavelength context to best understand the
environment of VRO 42.05.01.
We used the 1420 MHz continuum maps from the Canadian Galactic Plane Suvey \cite[CGPS,][]{taylor03} to show the regions of synchrotron emission.
We also used the 21 cm data cubes from the CGPS to study the neutral hydrogen distribution around the remnant. 
In the infrared we used data from the Wide-Field Infrared Survey Explorer all-sky survey \cite[WISE; ][]{wright10},
at wavelengths of 3.4~$\mu$m, 4.6~$\mu$m, 12~$\mu$m, and 22~$\mu$m.
Finally, we took the optical H$\alpha$ and [S II] maps from the Middlebury Emission-Line Atlas of Galactic SNRs\footnote{ 
\protect\url{http://sites.middlebury.edu/snratlas/g166-04-3-vro-42-05-01/}} (Winkler et al., in preparation). 
The optical images were taken with the 0.9~m KP Schmidt telescope at the Kitt Peak National Observatory between January 
15 and 19, 1996. 
The H$\alpha$ filter had a central wavelength of 6560.9~\angstrom\ and a width of 25.3~\angstrom, and the [S II] filter
was centred at	6718.3 \angstrom\ and was 47.9 \angstrom\ wide.

\section{Results}

In the region of our observations, for all three CO lines, there are two velocity ranges where emission is present. 
Between roughly $-28$ and $-12$ km\,s$^{-1}$ (which we refer to as the distant component)
different structures appear and fade continuously within the range, and, after a few dark channels, the emission
begins to reappear in the range of $-8$ to $-1$ km\,s$^{-1}$ (which we refer to as the close component).
The remaining velocity ranges included in our observations show only noise. 

\subsection{Method}

\begin{figure}
\includegraphics[width=\columnwidth]{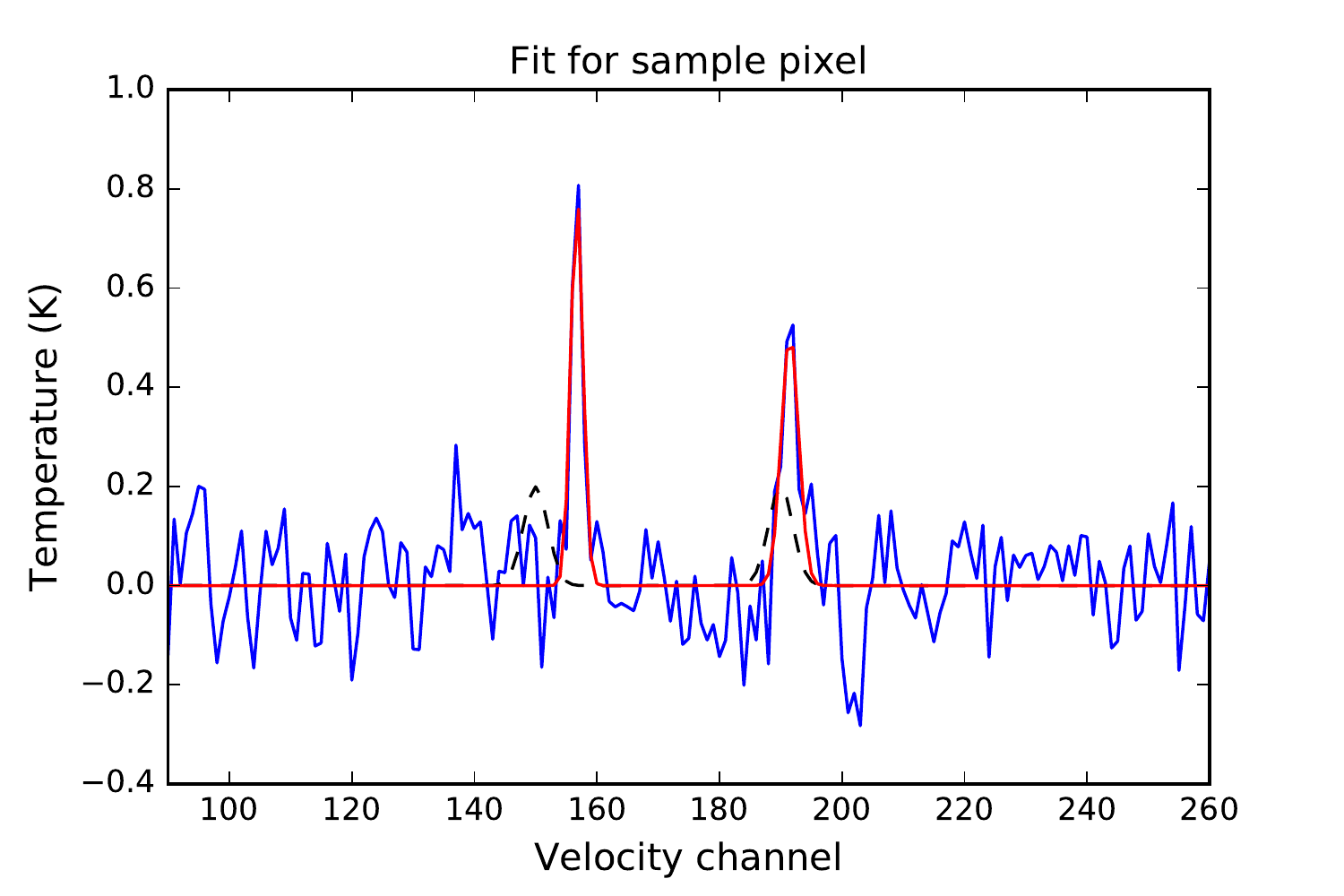}
      \caption{Example of the fit that produces Fig. \ref{results}
for a single pixel. The data points from the cube are plotted in blue, the initial guess for the
fit in dashed black, and the fit result is in red. The x-axis is in units of the cube velocity channels, where channel 200 corresponds
to 0~km\,s$^{-1}$, and channels are 0.5~km\,s$^{-1}$ apart. For this pixel the peaks are 0.785 K at $-21.6$~km~s$^{-1}$ (channel 156.7),
and at 0.508 K at $-4.2$~km~s$^{-1}$ (channel 191.5).}
         \label{sample_pix}
\end{figure}

\begin{figure*}{
         \centering
            \includegraphics[width=0.9\textwidth]{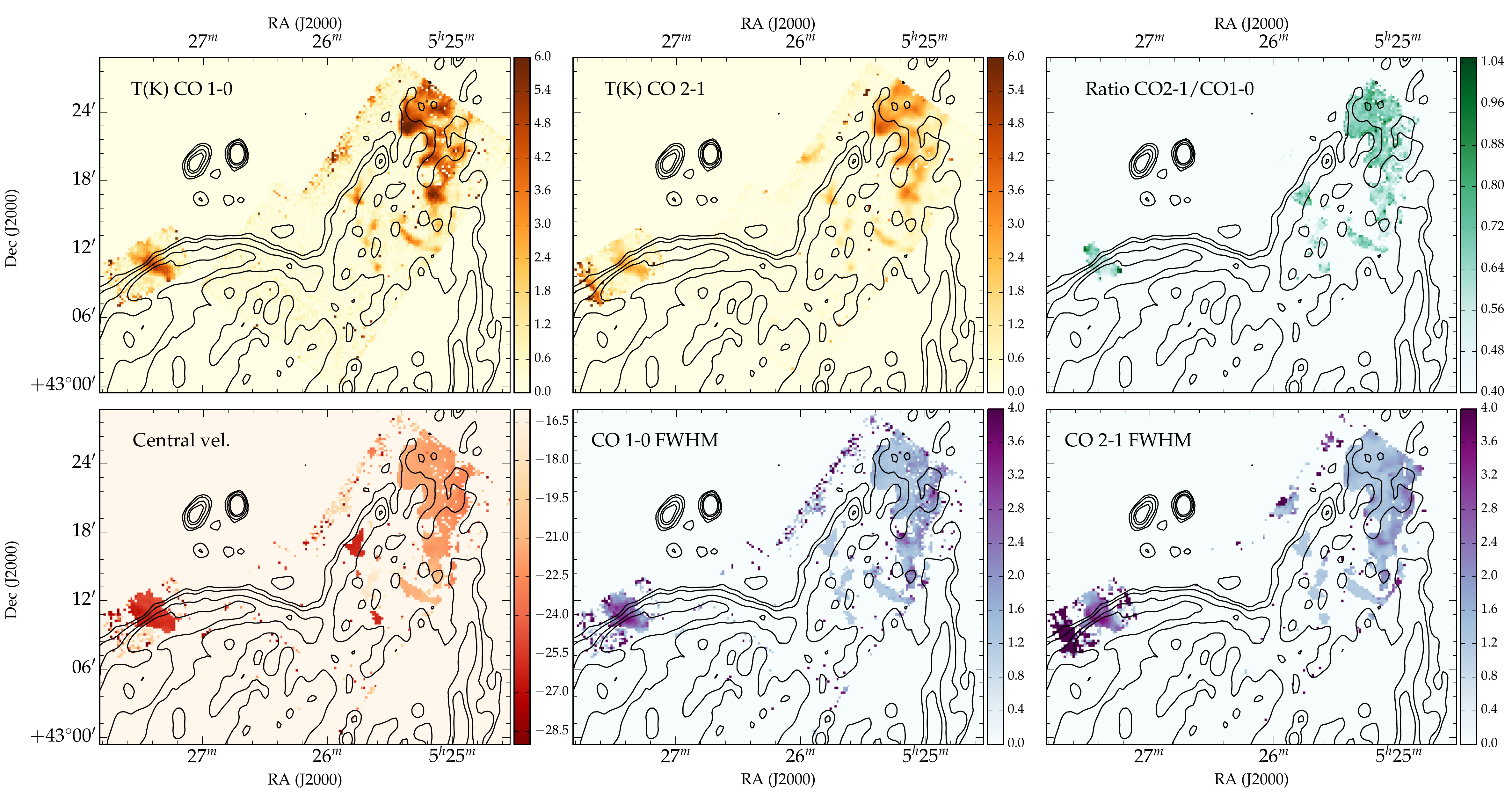}
            \includegraphics[width=0.9\textwidth]{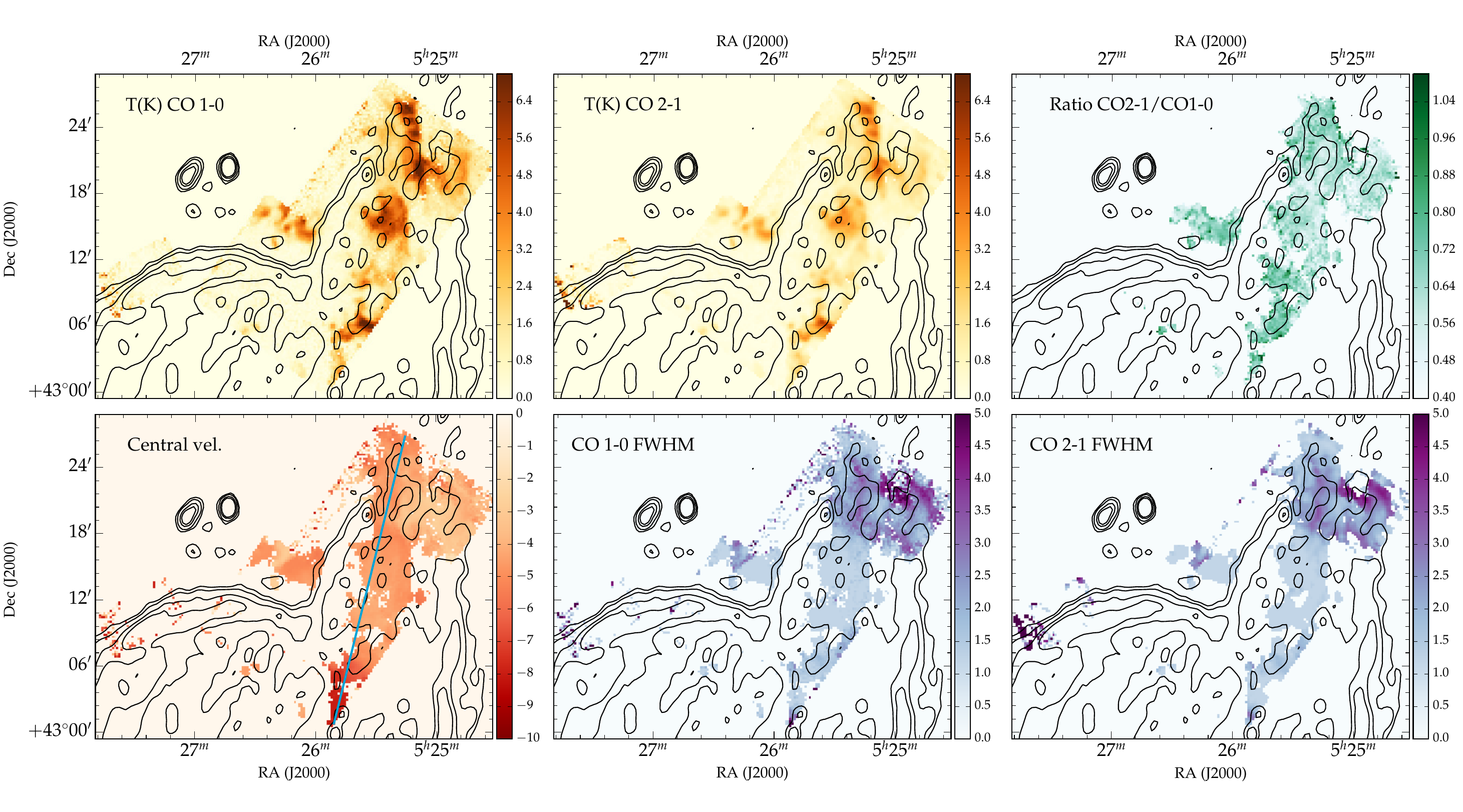}
      \caption{Results from fitting the distant (top) and close (bottom) components of the line profile to a Gaussian.
      For each of the two panels: height of the $^{12}$CO $J=1-0$ line in K (top, left),
      height of the $^{12}$CO $J=2-1$ line in K (top, middle), full-width at half maximum of the $^{12}$CO $J=1-0$ line
      in km~s$^{-1}$ (bottom, middle), 
      full-width at half maximum of the $^{12}$CO $J=2-1$ line km~s$^{-1}$ (bottom, right), and best-fit central velocity in km~s$^{-1}$ (bottom, left). For regions with signal over 5$\sigma$,
      the ratio of $^{12}$CO $J=2-1$ to $^{12}$CO $J=1-0$ is plotted in the top, right. Overlaid are the radio contours at 1420 MHz
      in units of brightness temperature at 4.8, 5.0, 5.5, and 6.0 K.
      Note the velocity gradient
      from the south to the north-west in the best-fit central velocity of the close component plot (lower left corner). 
      The line used to make the position-velocity plot in Fig. \ref{posvelslice} is indicated in this map
      in dark blue.
      }
         \label{results}}
\end{figure*}

We first aimed to determine some of the 
basic parameters of the molecular clouds observed in our data. 
In order to measure the peak temperature ($T_0$), velocity spread ($\Delta v$), and central velocity of the emission components, 
we fitted each pixel in the data cube
to two gaussians, one restricted to the velocity range of the distant component, and one restricted to the range of the close component. 
For each gaussian the fit produced a height (peak temperature), full-width at half-maximum (velocity spread), and centre (velocity
of the peak). 
A sample of what the fit did for each pixel is shown in Fig. \ref{sample_pix}.
The results for the $^{12}$CO transitions are plotted in Fig. \ref{results}. 

\subsection{Ratio maps}

Ambient molecular clouds have a temperature of $\sim10$~K, typical in the ISM. For 10~K, the ratio
of $^{12}$CO $J=2-1$ and $^{12}$CO $J=1-0$ is smaller than one. In shock-heated gas, the temperature
rises to $T\gg10$~K, which causes the $^{12}$CO $J=2-1$ and $^{12}$CO $J=1-0$ ratio to become 
larger than one (provided that the lines are optically thin), or approximately one (in the optically thick case). For this reason,
the ratio of $^{12}$CO $J=2-1$ to $^{12}$CO $J=1-0$ is commonly used as an indicator of the presence of shocks in a 
molecular cloud \citep{seta98}.

For the pixels where the $^{12}$CO $J=2-1$ and $^{12}$CO $J=1-0$ best-fit velocities of the peaks
matched, and for which the best-fit peak temperature was more than five times above the rms noise for that pixel (Fig. \ref{rms_noise}),
we took the ratio of the peak temperatures. These are plotted in the top-right panels of Fig. \ref{results}.
All the values are below one, which indicates that these are ambient molecular clouds that have not
been heated by the SNR shock.

\subsection{Column density maps}

It is possible to estimate physical parameters such as optical depth and molecular column density
from the $^{13}$CO $J=1-0$ measurements, provided that the rotational levels of the $^{13}$CO molecule
are in local thermodynamic equilibrium (LTE). 

In practical units, the optical depth and column density of $^{13}$CO are \citep{kawamura98}:
\begin{equation}
\tau_0^{13} = - \ln \left[ 1 - \frac{T_0^{13}}{\frac{5.29}{e^{5.29/T_\mathrm{ex}}-1}-0.87} \right],
\end{equation}
and
\begin{equation}
\left[\frac{N^{13}}{\mathrm{cm}^{-2}}\right] = 2.42\times10^{14} \left(\frac{\Delta v}{\mathrm{km\,s}^{-1}} \right) 
\left(\frac{T_\mathrm{ex}}{\mathrm{K}}\right) \frac{\tau_0^{13}}{1 - e^{-5.29/T_\mathrm{ex}}}.
\label{columndensity}
\end{equation}
Here, 
\begin{equation}
T_0^{13} = \left[J^{13}(T_\mathrm{ex}) - J^{13}(T_\mathrm{bg})\right] \left(1 - e^{-\tau_0^{13}}\right),
\end{equation}
where $T_\mathrm{ex}$ is the excitation temperature of the $J=1-0$ transition of $^{12}$CO in K (the same as for  $^{13}$CO), 
$T_\mathrm{bg} = 2.7$ K, and $J(T) = 1/(\exp[-5.29/T] - 1)$.

We took the excitation temperature of the $J=1-0$ transition of $^{12}$CO to be $T_\mathrm{ex}=10$~K.
We cannot calculate the excitation temperature directly from our measurements because we would need to assume
that the $J=1-0$ transition of $^{12}$CO is optically thick everywhere (this assumption can break down at the edge of the clould,
which could fall within our pixel) and also that the filling factor $f$ is 1.
The low $T_\mathrm{mb}$ temperatures we measured (see Figs. \ref{molec_data_dist}, \ref{molec_data_local})
suggest that the molecular cloud might be composed of clumps smaller than the telescope beam ($f<1$).

\begin{figure*}
\includegraphics[width=\textwidth]{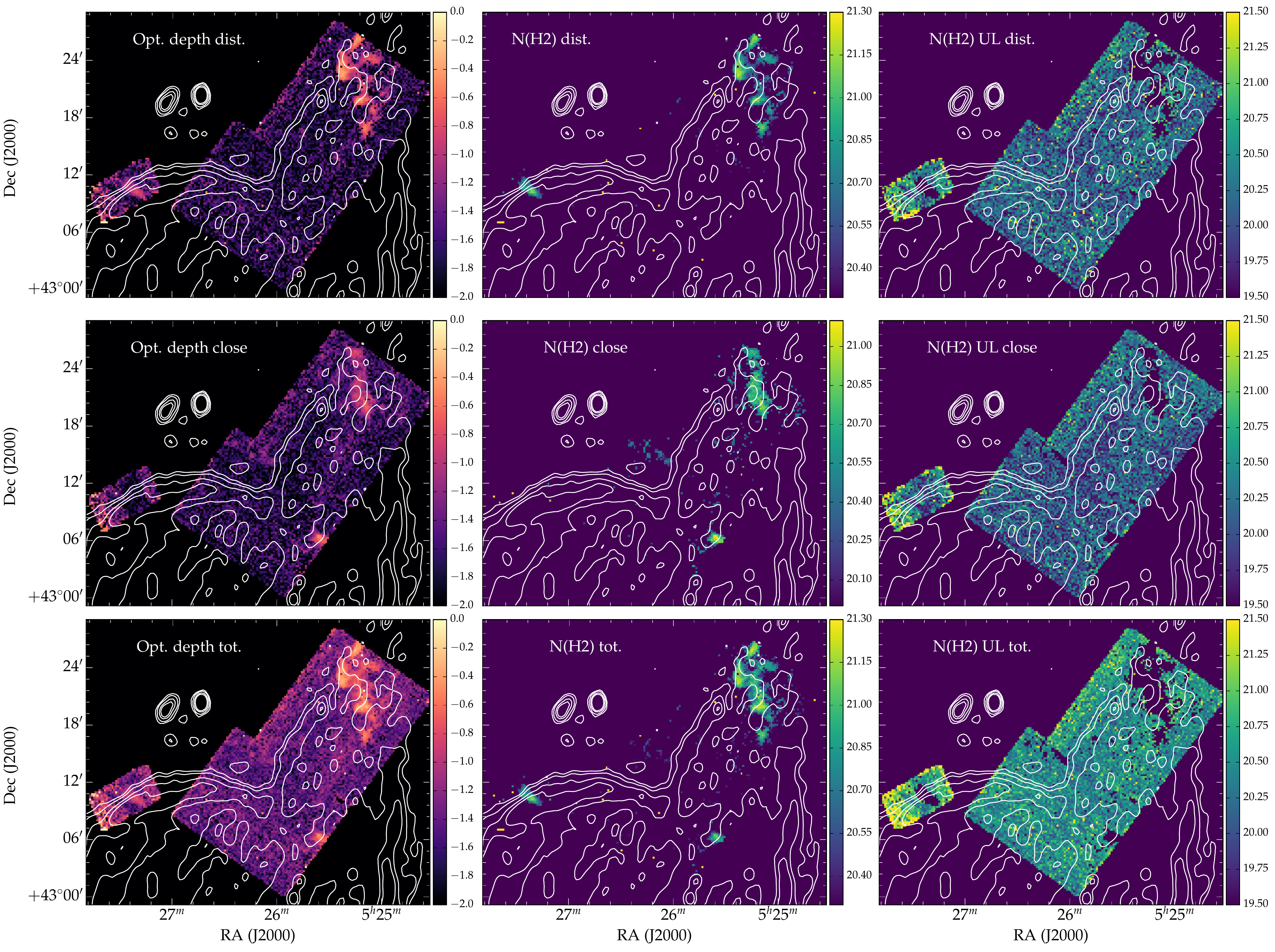}
      \caption{From left to right, log of the optical depth of $^{13}$CO ($\log \tau_0^{13}$), log of the $3\sigma$ column density of H$_2$ in cm$^{-2}$, and
      log of the upper limit in the column density of H$_2$ in cm$^{-2}$. Overlaid are the radio contours at 1420 MHz
      in units of brightness temperature at 4.8, 5.0, 5.5, and 6.0 K.
      The difference between the middle and right columns is that for the middle column the $N(\mathrm{H}_2)$ was calculated for the regions where the peak temperature of
      $^{13}$CO was 3 times or more the rms value for that pixel (i.e. right panel in Fig. \ref{rms_noise}), whereas for the right column the peak $^{13}$CO was within $3\sigma$,
      resulting in only an upper limit.
      The top panels correspond to the distant component ($-27.5 \, \mathrm{~km\,s}^{-1}$ to $-15.5 \, \mathrm{~km\,s}^{-1}$), the middle panels correspond
      to the close component($-8.5 \, \mathrm{km\,s}^{-1}$ to $-1.0 \, \mathrm{km\,s}^{-1}$), and the bottom
      panels correspond to the entire velocity space of our observations.}
         \label{columndens_tot}
\end{figure*}

Using equation \ref{columndensity}, 
$T_\mathrm{ex}=10$~K, a $N($H$_2)$ to $N(\mathrm{CO})$ ratio of $5 \times 10^5$ \citep{dickman78},
and assuming that the $^{13}$CO is in LTE, we obtained the maps in Fig. \ref{columndens_tot}. Due to the uneven noise levels in
our map (clearly visible in the rms plots in Fig. \ref{rms_noise}), we present the regions where we have a 3$\sigma$
detection of the hydrogen column density and those for which we only have an upper limit separately.

\subsection{Position-velocity maps for possible wind component}

One can see from the plot of best-fit central velocity of the close component (Fig. \ref{results}, lower-left corner) 
that there is a velocity gradient from the position RA=5:25:50, Dec=+43:00:34 that goes north-west, at an angle
of 72$^\mathrm{o}$ with the horizontal. 

\begin{figure*}
\includegraphics[width=\textwidth]{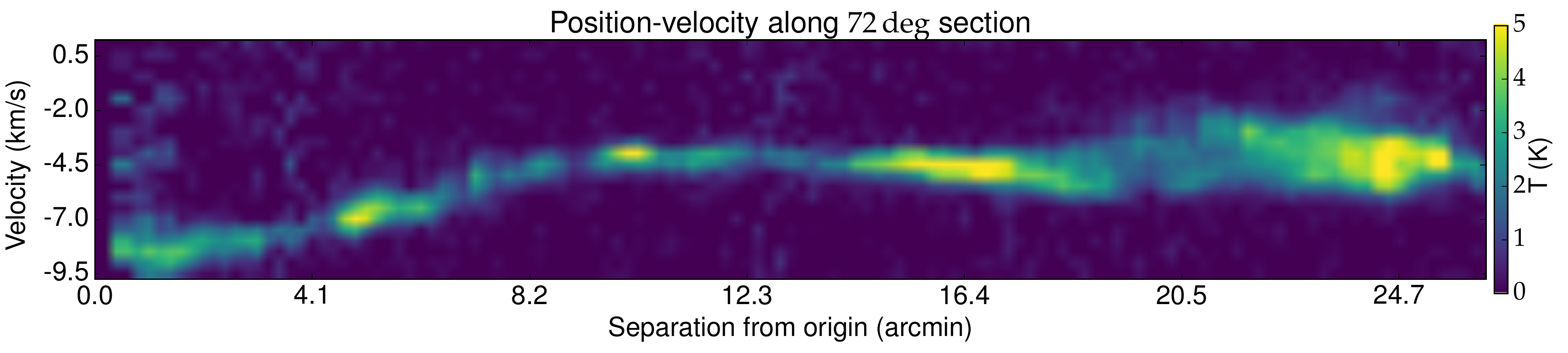}
      \caption{Position-velocity plot for a line that goes from RA=5:25:50, Dec=+43:00:34 to
RA=5:25:09, Dec=+43:25:36 with a 72$^\mathrm{o}$ inclination (see Fig. \ref{results}, lower-left corner) along the large velocity gradient
      seen at velocities of $-8$ km\,s$^{-1}$ to $-2$ km\,s$^{-1}$.}
         \label{posvelslice}
\end{figure*}

Later in this paper we argue that this gradient is a result of the interaction between a stellar wind and the molecular cloud.
In Fig. \ref{posvelslice} we present a position-velocity diagram of a slice taken along the line that goes from RA=5:25:50, Dec=+43:00:34 to
RA=5:25:09, Dec=+43:25:36, forming a 72 degree angle with the horizontal, as indicated by the blue line in Fig. \ref{results} lower-left map. 

It is clear from the position-velocity diagram that there is a part of the cloud (near RA=5:25:50, Dec=+43:00:34)
that is moving differently from the quiescent velocity of the system of $-4.5$~km~s$^{-1}$. The measured velocity 
difference is of $4.5$~km~s$^{-1}$ in the blue-shifted direction, although the real velocity difference could be larger,
since this feature is visible in the edge of our map. One can see
in the top four maps of Fig. \ref{molec_data_local}
that this disturbed component begins to appear in the southern edge of our map and moves north-west.

\section{The molecular environment of VRO 42.05.01}

\subsection{A clumpy molecular medium and a diffuse environment}

The channel maps show that the molecular gas near VRO 42.05.01 is very clumpy. The $^{12}$CO $J=1-0$ emission
has a main beam temperature $T_\mathrm{mb}\lesssim7$~K for the vast majority of pixels (see Fig. \ref{results}, and channel maps in
Figs. \ref{molec_data_dist} and \ref{molec_data_local}), cooler than the $\sim10$~K typical of ISM molecular clouds.
This points to the presence of unresolved clumps, of angular size smaller than the telescope beam ($\theta_\mathrm{clump}<0.11$~pc for a distance of 1~kpc, for
the close component; $\theta_\mathrm{clump}<0.68$~pc for a distance of 6~kpc, for
the distant component).

Another important aspect to note is that the SNR is in an environment with a low molecular column density (see $3\sigma$ measurements and upper limits
in the $n_\mathrm{H_2}$ distribution in Fig. \ref{results}). Our measured column densities are
as high as $1.7\times10^{21}$~cm$^{-2}$ for only three small ($\sim1\arcmin$) regions in the face of the SNR, and for most of the remnant
we measure an upper limit of $3.1\times10^{20}$~cm$^{-2}$ or lower. These values are
smaller than earlier X-ray measurements
in \cite{guo97}, who found $\sim~2.9\times10^{21}$~cm$^{-2}$ for regions of $11'\times11'$, although their X-ray observations and our
IRAM observations do not overlap. Moreover, the X-ray measurement of column density is sensitive to neutral and ionised hydrogen, in addition to
molecular hydrogen, so it is not surprising that the column density measured here is smaller.

\subsection{There is no physical proof of interaction between the supernova remnant shock and the molecular clouds in the region of
our observations}

\begin{figure}
\includegraphics[width=\columnwidth]{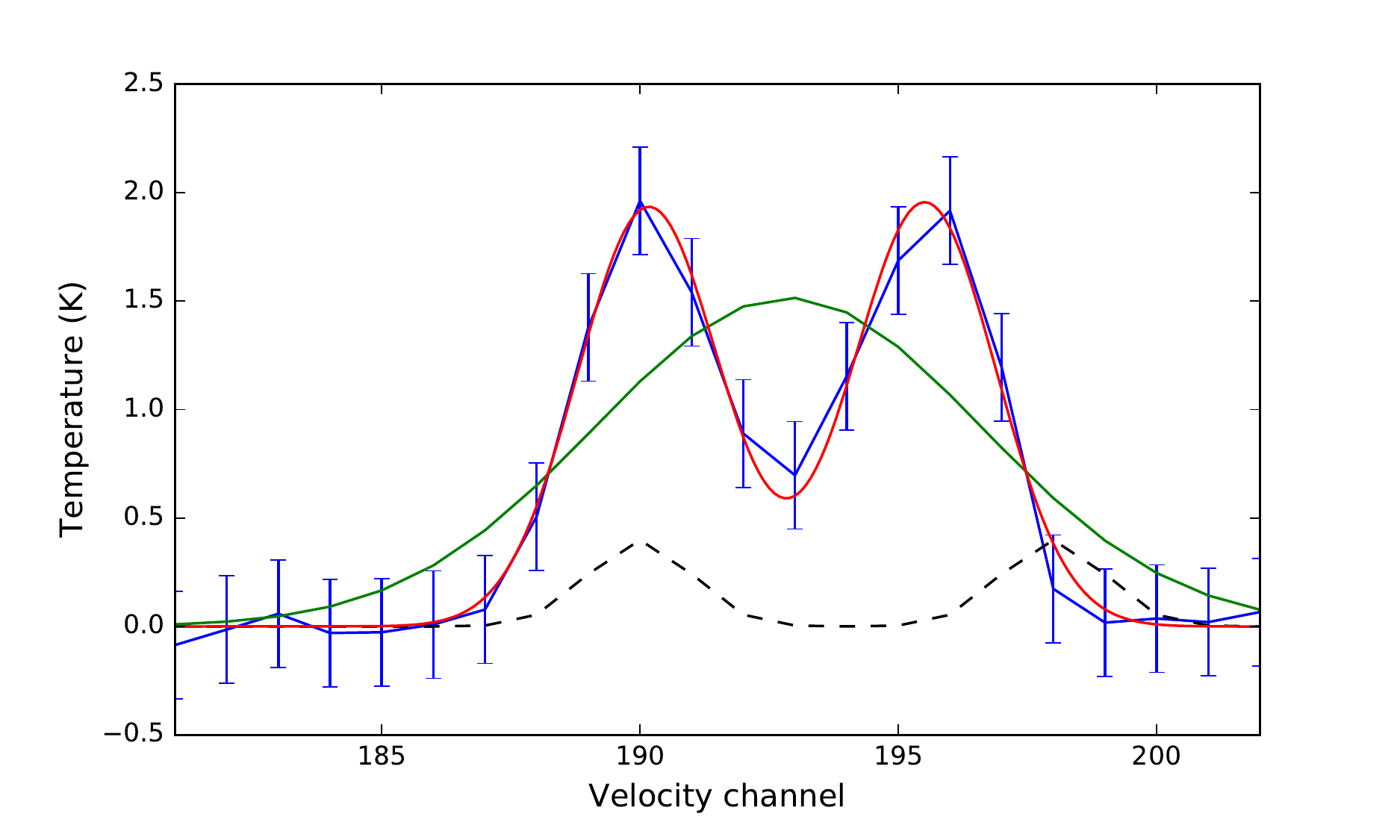}
      \caption{Averaged spectrum for the values in Fig. \ref{results}, blue plots, where the FWHM of the lines are greater than 3.5~km~s$^{-1}$.
      The fitting routine did not account for two clumps along the line-of-sight of a pixel for this range in velocities, and so tried to fit the two components as a single
      Gaussian with a broad FWHM (in green, corresponding to a FWHM of 4.4~km~s$^{-1}$). When we allowed for a second Gaussian
      term to be added to the model,  the two clumps are fitted by Gaussians with narrow FWHM (1.62~km~s$^{-1}$).
      The data points from the cube are plotted in blue, the initial guess for the
fit in dashed black, and the fit result is in red. The x-axis is in units of the cube velocity channels, where channel 200 corresponds
to 0~km\,s$^{-1}$, and channels are 0.5~km\,s$^{-1}$ apart.}
         \label{widewings}
\end{figure}

In our observations,
we do not find any conclusive physical proof that the SNR shock has reached its neighbouring molecular clouds. 
For the most part, we do not find broadened or asymmetric lines (see Fig. \ref{results}, in blue, for best-fit line widths), 
nor high ratios between lines of different excitation states (Fig. \ref{results}, in green).
The blue maps in Fig. \ref{results} (the bottom ones, of the close component) do show a region towards the west where the line widths are $>4$~km~s$^{-1}$.
Upon examination, these broad lines turn out to be an artefact of our fitting routine (see Fig. \ref{widewings}). There are two clumps
along the line-of-sight of these pixels, which the fitting routine tried to fit as one single Gaussian in this velocity range, resulting in a wider FWHM
and a shorter height. When we corrected by adding a second Gaussian component, the two clumps were fitted by Gaussians 
with narrow FWHM (1.62~km~s$^{-1}$), again indicating an unperturbed medium.

At an even more basic level, there are no obvious morphological coincidences between the supernova remnant 
and the molecular cloud, in the form of indentations or borders where these two structures encircle each other. The 
molecular cloud features that we see are on the face of the remnant, where the radio and optical 
emission are relatively faint (although this is likely because we are looking at the shock front edge-on and hence there is no limb-brightening,
and not because there is no shock there). 

We cannot say, from the IRAM observations, that the supernova remnant shock has reached and heated
its surrounding molecular cloud in the $26'\times14'$ region towards the west and the
$8'\times4'$ region towards the north of VRO 42.05.01 that we observed. The environment appears to be relatively empty of molecular clouds, 
we do not detect any sharp density gradients in the face of the remnant separating shell and wing, and,
at least in the region of our observations,
it is not likely that dense molecular gas is responsible for the strange shape of this source.

\subsection{Wind parameters}

\begin{figure}
\includegraphics[width=\columnwidth]{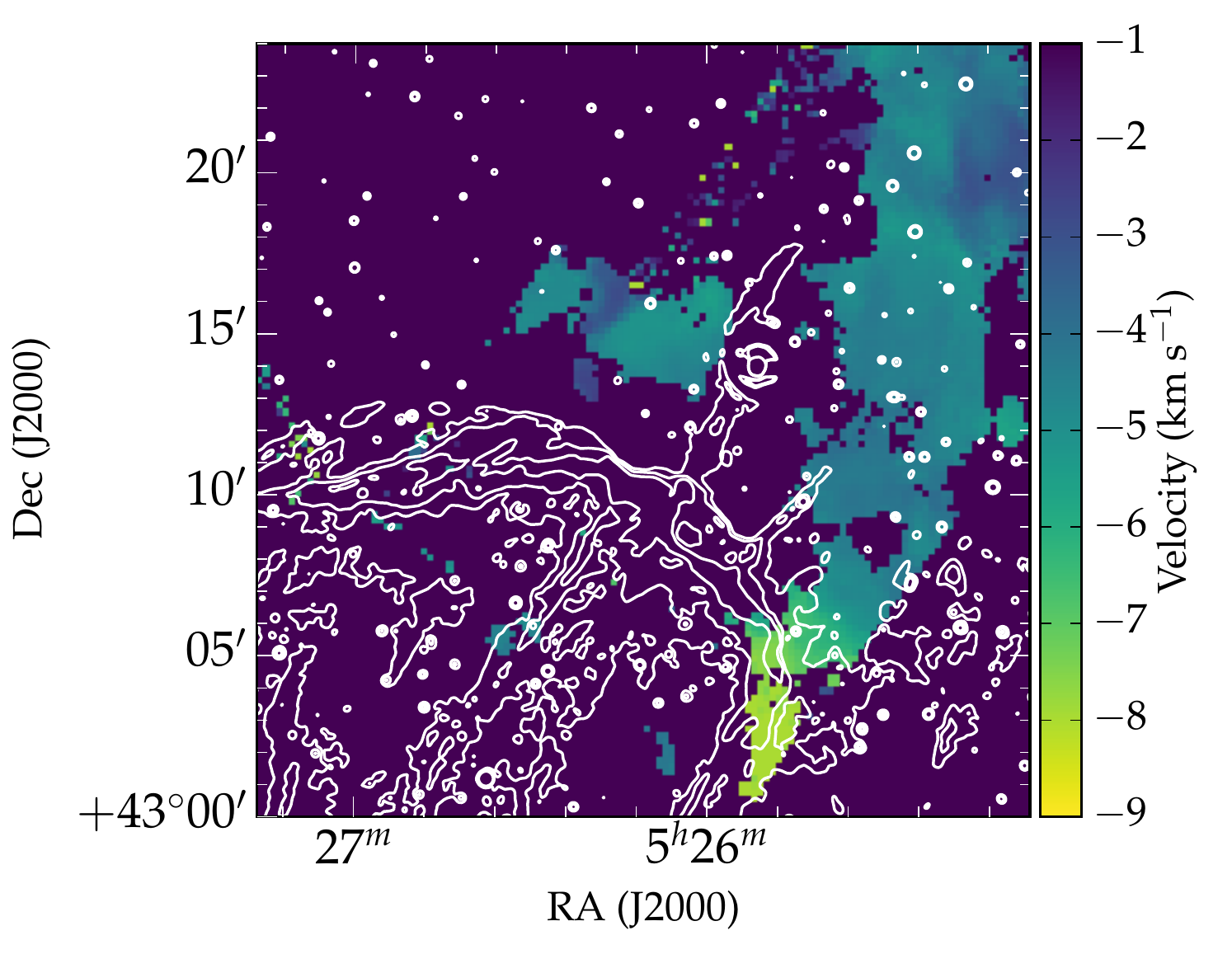}
      \caption{Best-fit central velocity for the close component (same as Fig. \ref{results}, lower left). Overlaid are the contours of VRO 42.05.01 in H$\alpha$
      at $(5, 10, 20)\times10^{-17}$~ergs~cm$^{-2}$~s$^{-1}$. 
      It should be noted that the velocity gradient (which fades from yellow to green) appears to follow the semicircle (bright in H$\alpha$, and also in radio synchrotron)
      of the small shell of VRO 42.05.01.}
         \label{optical}
\end{figure}

We can very roughly estimate some properties of the stellar wind from the expanding bubble revealed in our observations
(Fig. \ref{posvelslice}).

The detection of the velocity gradient is limited to a small region in the southernmost corner of our map. 
In fact, the region with a gradient appears to coincide with some features that follow the semicircle of the small
shell. The rim of the shell is more easily visible in the optical, as one can see in Fig. \ref{optical}. Although the
feature is, unfortunately, at the edge of our map, it seems to follow the H$\alpha$ morphology. For this reason we took
the radius of the wind $R$ to be the radius of the shell (which subtends an angle of 29\arcmin), 8.4~pc at a distance of 
1~kpc. 

Combining equations 51 and 52 in \cite{weaver77}, which describe the evolution of an interstellar bubble, we arrived
to:
\begin{equation}
L_\mathrm{w} = 2.2\times10^{33} \left(\frac{n_0}{1~\mathrm{cm}^{-3}}\right) \left(\frac{R}{16.9~\mathrm{pc}}\right)^2  \left(\frac{v}{4.5~\mathrm{km~s^{-1}}}\right)^3 \mathrm{erg~s}^{-1}
\end{equation}
where $L_\mathrm{w}$ is the luminosity of the stellar wind, $v$ is the expansion velocity of the shell (the difference between the velocity of the molecular cloud and the disturbed
component, as one can see in Fig. \ref{posvelslice}; we take $v=4.5~\mathrm{km~s}^{-1}$, although this is a lower limit),
$n_0$ is the number density of hydrogen atoms, and $R$ is the radius of the bubble. 
\cite{pineault87} propose that the density of the medium that surrounds the SNR could be
1~cm$^{-3}$. We use this density value for our mass-loss estimate, but keep the dependency
on $n_0$ explicit.

The wind luminosity is related to the mass-loss
rate of the star as:
\begin{equation}
L_\mathrm{w} = \frac{1}{2}\dot{M}v_\mathrm{w}^2.
\end{equation}
where $v_\mathrm{w}$ is the wind velocity and $\dot{M}$ is the mass-loss rate of the star due to the wind.

A luminosity $L_\mathrm{w} = 2.2\,n_0\times10^{33}$~erg~s$^{-1}$, and a wind velocity of 1000~km~s$^{-1}$
corresponds to a mass-loss rate of $7.7\,n_0\times10^{-9}$~\msun~yr$^{-1}$, over a period of $1.1\times10^6$~yr.
These are reasonable parameters for the main-sequence wind of a star of $12-14$~\msun\ \citep{chen13}, which could be the progenitor of VRO 42.05.01.

\subsection{Geometry of the system}

\begin{figure}
\includegraphics[width=\columnwidth]{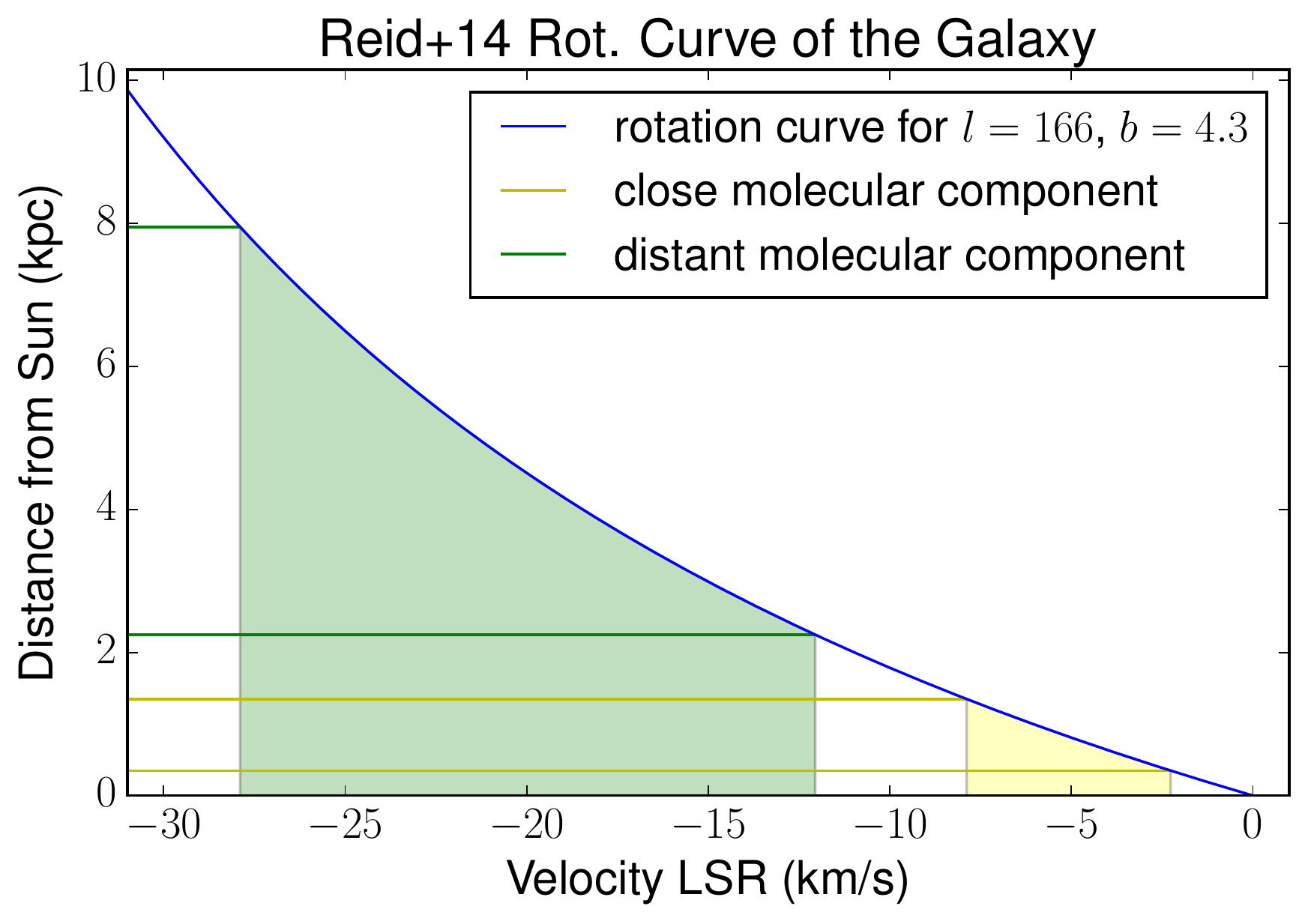}
      \caption{Distance from the Sun as a function of measured local standard of rest velocity as per \cite{reid15}.
      The distant molecular component corresponds to distances between 2.25 and 8 kpc, and the close component
      corresponds to distances between 0.35 and 1.35 kpc.}
         \label{rot_curve}
\end{figure}

\cite{landecker89} first studied the neutral hydrogen environment of VRO 42.05.01. They identified a series of features
that seemed to indicate that the SNR was interacting with a surrounding H I structure at a systemic
velocity of $-34$ km\,s$^{-1}$. In particular, they found a cavity in the H I distribution
at $-28$ km\,s$^{-1}$, which, they proposed, is an interstellar cavity in which the remnant is located. 
Due to the large errors in kinematic distances in the direction of the Galactic anti-centre, they could only estimate that the distance to the cavity
(and, hence, the remnant) should be the distance of the Perseus arm, which they take to be $4.5\pm1.5$~kpc.

In our earlier work on VRO 42.05.01
\citep{arias19}, we followed this claim up by examining the H I cubes from the CGPS \citep{taylor03}. We did not see some of the
purported features of interaction listed in \cite{landecker89}, including the alleged H I cavity.
Initially we were inclined to believe that the systemic velocity of VRO 42.05.01
was $-28$ km\,s$^{-1}$, that of the cavity, or otherwise close to the velocity of the H I structure
that \cite{landecker89} associated with the remnant, $-34$ km\,s$^{-1}$, but
 the observed distribution in molecular gas points to the remnant being significantly closer.

To convert velocities to distances, we used the rotation curve of the outer Galaxy in
\citet[][plotted for the coordinates of VRO 42.05.01 in Fig. \ref{rot_curve}]{reid15}.
According to this curve, the component
we refer to as \lq close' is located between 0.35 and 1.35~kpc from the Sun, and the \lq distant' component is between
2.25 and 8~kpc away. From the central velocity plots (Fig. \ref{results}, in red) we see that the bulk of central velocities in the 
close component are approximately $-4$ km\,s$^{-1}$, and for the distant component, $-20$ km\,s$^{-1}$.
This corresponds to $65\pm5$~pc and  $4.6\pm0.5$~kpc, respectively.  
We expect the SNR to be further away than $-4$ km\,s$^{-1}$, since the region with a velocity
gradient has not been shocked, but close enough to that velocity that its progenitor could blow the wind bubble
that caused the gradient. This suggests that VRO 42.05.01 has a local standard of rest velocity of approximately $-6$ km\,s$^{-1}$, corresponding to a distance of $1.0\pm0.4$~kpc.
We note, however, that the flat rotation curve does not always apply to the Perseus arm
\cite[e.g.][]{roberts72,foster15}, and so these distance estimates should be taken with a grain of salt.

If the blue-shifted velocity gradient observed in Fig. \ref{posvelslice} is indeed due to a wind blown by the 
progenitor star of the remnant, then the remnant should be behind the cloud. It should be far enough from the cloud that the remnant
shock front has not yet reached and heated the molecular cloud (as evidenced from the $^{12}$CO $J=2-1$ to $^{12}$CO $J=1-0$
ratio, Fig. \ref{results}), but still close enough that the companion wind could have reached and disrupted the molecular gas.
Due to the large distance ($>1$~kpc) between the close and distant molecular components, according to our interpretation, the supernova
remnant and the distant velocity component are unrelated. This distant component is probably
associated with the Perseus arm.

If VRO 42.05.01 was indeed 4.5 kpc away, at a Galactic latitude of 4.3$^\mathrm{o}$, this would correspond
to 340 pc above the Galactic plane, and the physical size of the remnant in its longer direction (45\arcmin) would be 59 pc. 
The height above the plane is slightly problematic if VRO 42.05.01 is the remnant of a core-collapse
explosion, because the scale height of the Milky Way thin disk (where massive stars typically reside) is $\sim300$~pc \citep{juric08},
and for OB associations the scale height is only $\sim40$~pc \citep{bobylev16}.
If, as we suggest, the distance to the remnant is $\sim1$~kpc, its height above the Galactic Plane would be 75 pc,
and its angular size would correspond to 13 pc.

According to our line of reasoning, we have the following scenario: the supernova remnant is outside the edge of the close molecular
cloud ($\sim1$~kpc away). The stellar wind has had to blow a cavity larger than the shock front of the remnant, at least in 
the direction of the molecular cloud (for the shell). For our rough estimate of the luminosity of the wind and mass-loss rate of the 
progenitor we took the radius of the bubble to be the radius of the small shell. In fact, the radius of the small shell delimits
the shock front; 
if the 
remnant is right at the edge of the close molecular cloud, the stellar wind would have had to blow a cavity slightly larger than 8.4~pc.

\section{The dust and H I environment of VRO 42.05.01}

\begin{figure*}
\includegraphics[width=\textwidth]{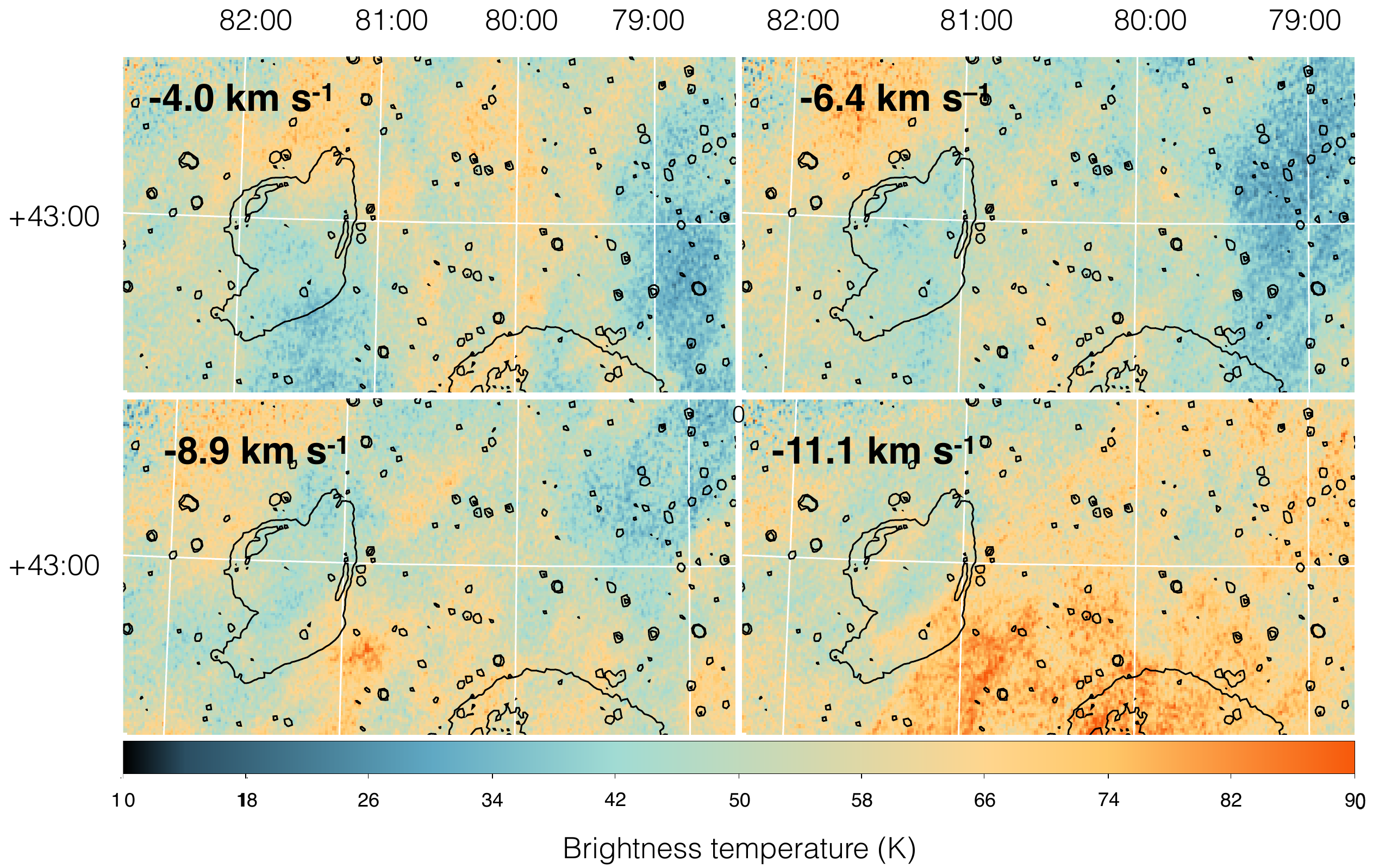}
      \caption{Velocity distribution of HI emission towards VRO 42.05.01 at the range of velocities that corresponds to the
      close molecular cloud. The data cube was taken from the CGPS. 
      Overlaid are the radio contours at 1420 MHz in units of brightness temperature at 2.5, 3.7, 4.8, and 6 K  .
      The wedges in the upper left corner of each plot correspond to the velocity of each slice. The brightness temperature $T_\mathrm{B}$ 
      colour scale is the same for all plots.}
         \label{neutral_hy}
\end{figure*}

\begin{figure*}
\includegraphics[width=0.9\textwidth]{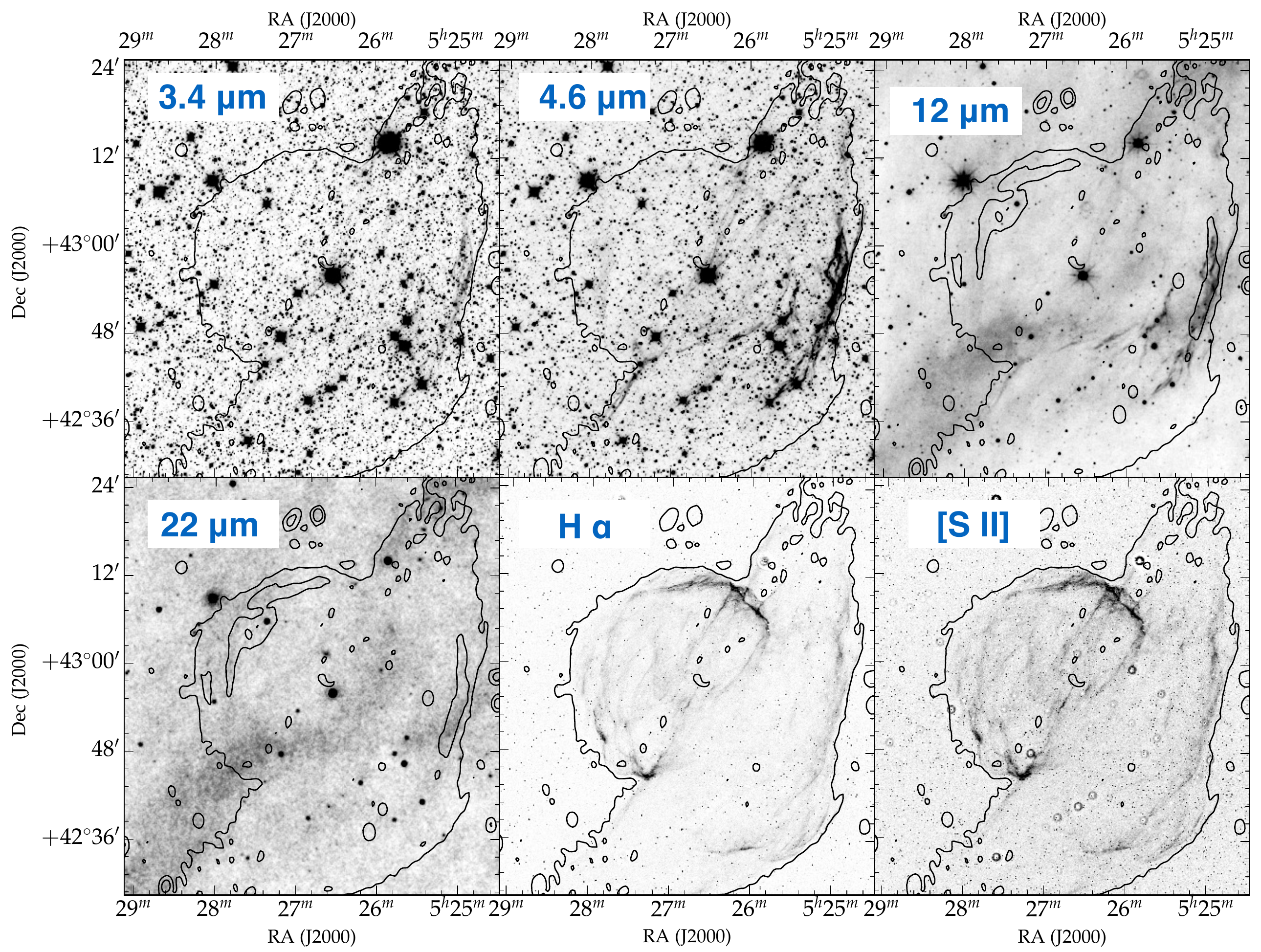}
      \caption{ 3.4~$\mu$m, 4.6~$\mu$m, 12~$\mu$m, and 22~$\mu$m \cite[WISE,][]{wright10}, 
H$\alpha$ and [S~II] (Middlebury Emission-Line Atlas of Galactic SNRs, Winkler et al., in preparation) maps of VRO 42.05.01.
Overlaid are the radio contours at 1420 MHz in units of brightness temperature at 4.8 and 6.0 K.}
         \label{IRopt}
\end{figure*}

\subsection{An H I cavity at $-6$~km~s$^{-1}$}

In light of our new suggested distance for VRO 42.05.01, we revisited the H I cubes from the CGPS for features
at velocities closer than those around the range of $-22$ km\,s$^{-1}$ to $-54$ km\,s$^{-1}$studied by \cite{landecker89}, and that we followed up on in \cite{arias19}.
The H I environment at the velocity ranges around our alleged systemic velocity for VRO 41.05.01, $-6$ km\,s$^{-1}$,
is quite crowded. The most suggestive features we found in the CGPS cube are
a cavity with roughly the shape of the SNR, which can be seen in Fig. \ref{neutral_hy} at $-6.4$ km\,s$^{-1}$, and an H I bright 
region just north of the remnant, in the shell region (which has traditionally been thought to be in a dense environment).
However, the H I data shows a lot of structure,
and coincidental similarities between the remnant and H I features are possible. A detailed study of the H I features at these velocities,
such as the one done by \cite{landecker89}, is beyond the scope of this paper, but for completeness we wanted to note that the 
H I data is not at odds with our suggested value for the systemic velocity as derived from the CO observations.

The H I bright region near the shell, north of the remnant (seen most clearly in Fig. \ref{neutral_hy}, second panel, at $-6.4$ km\,s$^{-1}$)
also coincides with a very bright radio and optical spot in VRO 42.05.01's small shell. In \cite{arias19} we note that this region has a radio spectral index that
steepens at low frequencies, which we argue is a result of the properties of high compression ratio shocks, and requires a high
post-shock density. 
If related, this bright H I feature could be responsible for the bright radio and optical emission, as well as
support our proposed explanation for low-frequency steepening.

\subsection{Hints from the infrared and the optical}

\begin{figure}
\includegraphics[width=0.9\columnwidth]{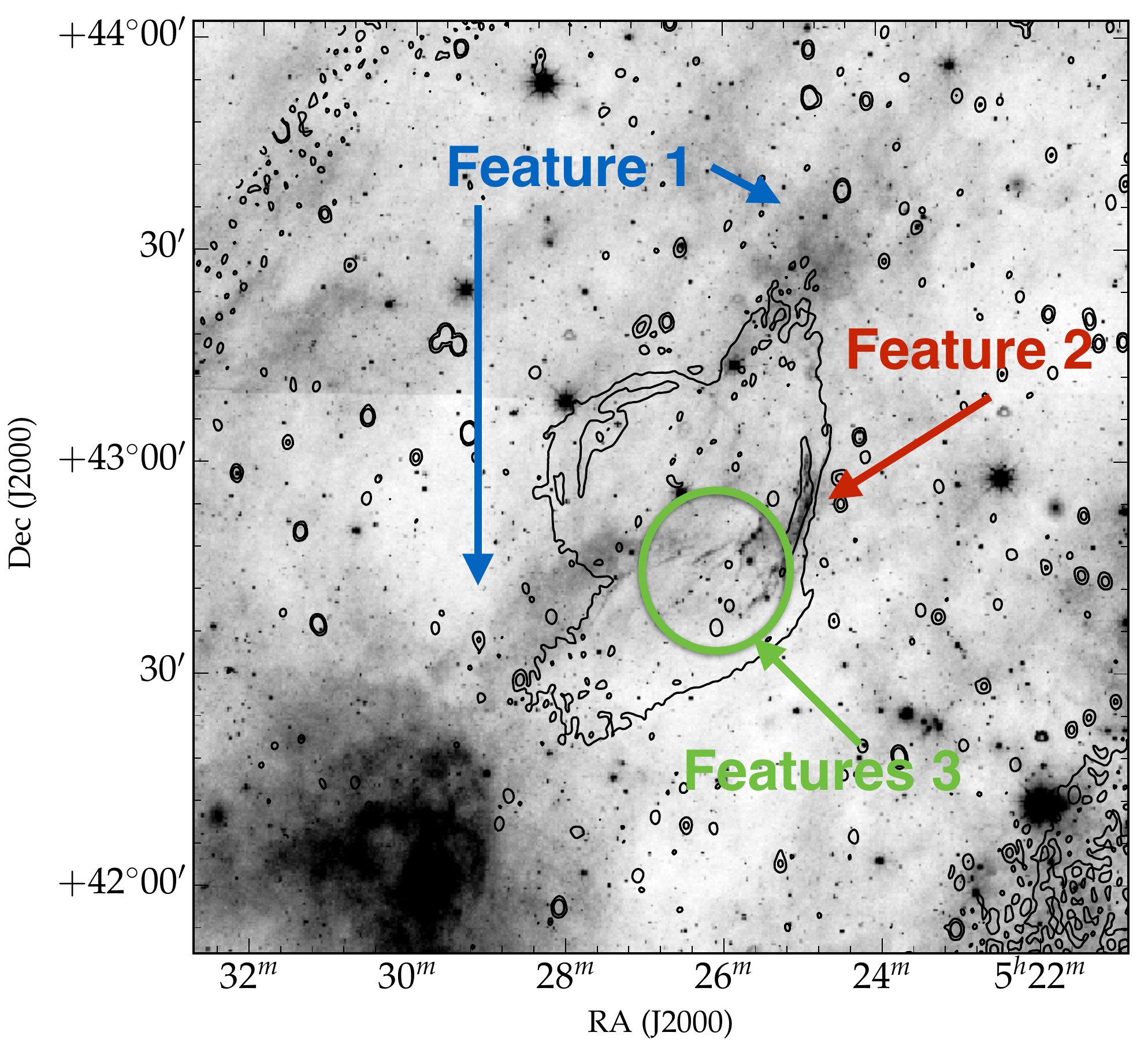}
      \caption{12~$\mu$m WISE map with 1420 MHz radio contours in units of brightness temperature at 4.8 and 6.0 K. 
      Labelled are the features we refer to in the body of the text.}
         \label{IRlab}
\end{figure}

The infrared is an important regime for understanding the Galactic environment.
In particular, infrared emission can be a diagnostic of interaction of a SNR with relatively dense gas, 
since shocked gas cools through line emission, and many important cooling lines occur in this part of the spectrum
\citep{reach06}. 

In Fig. \ref{IRopt} we display maps of VRO 42.05.01 
at 3.4~$\mu$m, 4.6~$\mu$m, 12~$\mu$m, and 22~$\mu$m from the WISE survey \citep{wright10}, as well
as the H$\alpha$ and [S~II] maps from the Middlebury Emission-Line Atlas of Galactic SNRs (Winkler et al., in preparation). 
We note the following features (labelled for clarity in Fig. \ref{IRlab}):
\begin{enumerate}
\item The 12~$\mu$m and 22~$\mu$m maps show a broad band of emission that goes from the south-east corner of the map, along the line separating the shell and the wing,
and that seems to encircle the shell, continuing the morphology from the stellar wind we found in the IRAM molecular data, and that
goes on to follow the boundary between shell and wing to the north-western corner of the map. 
This feature is external to the remnant, and part of a large structure seen more clearly in Fig. \ref{IRlab} (labelled in blue).
In the region of our IRAM observations, this is the only structure that could conceivably be interacting with the SNR.
It could be the boundary of the density discontinuity invoked by models to explain the source.
We find it is the most likely candidate
to be responsible for the unusual shape of this SNR in the area covered by our IRAM observations.
\item There is a filament in the wing, to the west of the remnant, that goes from north to south, that matches a bright radio
synchrotron feature (labelled in red in Fig. \ref{IRlab}). 
This feature is visible in all the infrared bands, although it is very faint by 3.4~$\mu$m. It is also visible in the optical maps.
\item There are a series of filaments internal to the wing, seen at 4.6~$\mu$m and 12~$\mu$m, 
that have very faint radio and optical counterparts. These filaments broadly trace a semicircle
that continues the semicircle of the small shell.
\end{enumerate}

\noindent
\cite{reach06} note that infrared emission from SNRs can be due to synchrotron emission, shock-heated dust, atomic fine-structure lines,
and molecular lines, but that it is challenging to disentangle the contribution of each component from colour maps.

Infrared synchrotron emission from SNRs requires a high radio luminosity, and VRO 42.05.01 is a 
relatively faint object (7 Jy at 1 GHz for an angular size of 45\arcmin$\times$80\arcmin)
although with a flat radio spectral index, $\alpha \approx 0.37$ \cite[]{leahy05}.
The only feature that coincides with a radio-bright filament is feature 2.
However, the filament appears to be brighter at 12~$\mu$m than at 22~$\mu$m, which points to a different
emission mechanism than synchrotron.

Features 2 and 3 are similar in that they are filamentary, and likely associated with the radiative SNR  shocks. 
The main difference between them is that feature 2 coincides with radio- and optical-bright filaments, which clearly trace the
SNR shock, whereas the radio and optical emission associated with the filaments that compose feature 3 are very faint. 
Infrared atomic fine-structure lines, and vibrational lines from molecules are expected in radiative shocks: the dense gas quickly cools
to temperatures of $100-1000$~K and infrared transitions are excited \citep{reach00}.
There are a number of forbidden emission lines that fall within the WISE bandpasses: most notably, 
the $S(3)$ and $S(9)$ lines of molecular hydrogen (9.666~$\mu$m and 4.695~$\mu$m) fall within the 12~$\mu$m and the 4.6~$\mu$m
WISE bands. These lines have been detected in other MM SNRs \cite[W28, W44, 3C391;][]{reach00}.

The filaments that form feature 2 seem to radiate in the  3.4~$\mu$m, 4.6~$\mu$m, and 12~$\mu$m bands, although not in the 22~$\mu$m.
It is not possible to disentangle what elements or molecules are radiating in these bands without spectroscopic data. In the case of
 the filaments that compose feature 3, 
we suspect that the emission is due to molecular hydrogen, since it is visible at 4.6~$\mu$m and 12~$\mu$m, but not at 
3.4~$\mu$m and 22~$\mu$m. 
Feature 2 could also be due to interaction with molecular hydrogen, and the coincident 3.4~$\mu$m emission could be 
shocked dust.
This would signal interaction between the supernova remnant shock and a nearby molecular cloud; 
targeted molecular line observations towards this region could confirm this hypothesis.
Projection effects might be responsible for the faint radio synchrotron at the location of the filaments. A further point to note
is that the feature 3 filaments coincide with the peak in X-ray brightness as seen in Fig. 4 of \cite{bocchino09} and Fig. 1 of \cite{matsumura17}.

Feature 1 is part of a large sheet of dust that cuts through the remnant. Its coincidence with the flat line that
separates the shell and the wing, as well as its wrapping around the shell are very suggestive that
this structure has had a role in producing the two-shell shape of this source.
It is the only structure that we have found that has such a clear morphological similarity with the SNR
at large scales.

\section{Discussion}

In this section we gather the multi-wavelength information available on VRO 42.05.01, in order to understand in what ways
its environment is responsible for its shape, and for its mixed-morphology nature.
The data are starting to paint a more complete picture of the formation of this SNR, challenging
some of the models in the literature.
The observational evidence that we need to take into account is the
following:
\begin{itemize}
\item The molecular clouds in our IRAM observations reveal an inhomogenous medium
that has not been reached by the SNR shock, as evidenced from the  $^{12}$CO $J=2-1$ to $^{12}$CO $J=1-0$
ratio.
\item The IRAM data shows a velocity gradient in one of the molecular clouds, 
which suggests that the main-sequence wind of the progenitor of VRO 42.05.01 swept up the ambient cloud. This implies that
at this stage the wind bubble boundary is further away from the remnant centre than the SNR shock.
\item There is a broad band of infrared dust emission that follows the sharp line that separates the two distinct regions in the remnant (the shell
and the wind). 
\item There are hints of shocked H$_2$ in the central region of the remnant (feature 3; not covered by our IRAM observations). This is
indicative of SNR-MC interaction.
\item The H I emission from the CGPS shows a cavity roughly the shape of VRO 42.05.01 at around $-6$ km\,s$^{-1}$.
\item There is a large polarised bubble, likely a Faraday screen, whose boundary matches the straight line separating the
shell and the wing \citep{kothes04}.
\item In the radio, the remnant has a flat spectral index $\alpha\approx0.37$ \cite[]{leahy05}, but the spectrum steepens
at low frequencies \cite[$\sim150$~MHz,][]{arias19}.
\end{itemize}

\subsection{Implication for models that explain the morphology of VRO 42.05.01}

The explanation of the remnant's radio morphology dates back to \cite{pineault87}, who proposed that the SNR broke out from a warm
medium of intermediate density (the explosion site) into a hot, tenuous interstellar cavity formed by earlier SNe or stellar winds.
We elaborated on this model in \cite{arias19}, where we noted that if the star was moving supersonically at the time of the explosion,
the bow shock would have produced a triangular cavity of the shape of the wing, into which the remnant would have expanded.

We have found out that the environment of VRO 42.05.01 appears to be rather empty, except for scattered,
clumpy molecular clouds, and a dust band that cuts through the remnant.
The infrared band could have provided the density discontinuity (dust often forms sheet-like structures, and this band
could be a sheet in projection) that \cite{pineault87} invoked, although their model required a diffuse tunnel, as opposed to
a dense boundary. 

Suppose that the pre-SN star was indeed moving in a way that produced a Mach cone of 120$^\mathrm{o}$ angle, giving rise
to the wing. In this case the bow shock would (a) shock the ISM and (b) shock the stellar wind. 
We do not find any evidence of bow- or SNR-shocked molecular clouds, but the CO can have cooled in the intervening
time since the SN explosion, and hence it is possible that the molecular clouds were bow-shocked
but we do not see enhanced ratios today. (Alternatively, the velocity gradient could be due to the bow shock itself, 
rather than to the stellar wind shocking the gas.)
Our molecular observations have revealed that the progenitor star blew a main-sequence wind that has reached the neighbouring
molecular clouds, but that the shock front itself has not. It is possible that the high densities the SNR
shock has encountered (which make it so radiative, and steepen its radio spectrum at low frequencies) are
due to the bow-shocked stellar wind, rather than to the ISM.

A final note: a recent \textit{Suzaku} paper \citep{matsumura17} argued that VRO 42.05.01 
has recombining plasma in the shell (the smaller semicircle).
The authors proposed that this might be due to heat being conducted from the SNR to a molecular cloud that 
allegedly sits at the north of the remnant, which lowers the temperature of the SNR.
Our observations of the SNR shell are limited, but we can say that at least the small region
of the shell covered by our observations is molecular cloud-free.

\subsection{Implication for models that explain MM SNRs}

\cite{bocchino09} go into some discussion on the \cite{white91} and \cite{cox99} models for the formation of mixed-morphology
SNRs, and find neither can explain the observed metallicity, temperature, and density profile of the X-ray emission of VRO 42.05.01.
The model of \cite{white91} suggests that the way a SNR becomes mixed-morphology is by expanding into an inhomogeneous ISM
whose mass is mostly contained in small clouds. The blast wave overtakes these clouds, but they survive the shock passage, and are
slowly evaporating inside the SNR shell. \cite{bocchino09} calculated the parameters that describe the \cite{cox99} model for 
VRO 42.05.01, and find a lower limit in the ambient density of 10~cm$^{-3}$. From what we have learned about the environment of
VRO 42.05.01 in this work, it is more similar to the clumped medium required by the \cite{white91} model rather than the very dense
medium required by the \cite{cox99} model.

\cite{chen08} propose a different scenario for the formation of MM SNR Kesteven 27: the SN explosion happened within a cavity
caused by the progenitor star's winds and ionising radiation interacting with its dense environment. When the SN blast wave hit the cavity wall,
a reflected shock (different from the reverse shock) was produced, going inwards, and reheating the interior gas (a mix of ejecta
and swept-up circumstellar material). 

The reflected shock scenario \citep{chen08} is an interesting option for the formation of VRO 42.05.01. If the SN explosion happened in the wing,
as we proposed in \cite{arias19}, the shock could have been reflected by the walls of the Mach cone produced by the bow shock, and the reflected
shock could have reheated the swept-up material to X-ray emitting temperatures. Eventually the reflected shock would have found a second density
enhancement, that of the dust filament, resulting in the sharp straight radiative filaments that separate the shell and the wing. 

When it comes to explaining the centrally-peaked X-ray morphology of MM SNRs it is natural to think that 
a dense environment could trigger an early reverse shock, so that stellar
ejecta are heated to X-ray temperatures before expanding too far from the explosion centre. 
However, the reverse shock heats stellar ejecta, which would produce a metal-rich X-ray spectrum. Many MM SNRs
do in fact show metal abundances \citep{lazendic06b}, but \cite{bocchino09} found no evidence of enhanced metal abundances in VRO 42.05.01.
If the X-ray emitting material was heated by a reflected shock then it would be a mix of ejecta and ISM material, which is
more consistent with the \cite{bocchino09} findings.

\vspace{0.3cm}
\noindent
As we did in \cite{arias19}, we caution that these scenarios for the formation of VRO 42.05.01 are speculative, and that detailed 
hydrodynamical simulations are necessary to understand whether a SN explosion happening in a clumpy medium, next to a dust sheet,
and within the bow shock of a moving star could look the way VRO 42.05.01 does across the spectrum. 

%

\section{Conclusions}

In this work we observed a region in the direction of SNR VRO 42.05.01 with the EMIR receiver in the IRAM 30m telescope.
We mapped a $26'\times14'$ region towards the west of the SNR and a $8'\times4'$  region towards the north of the remnant
in the $^{12}$CO $J=1-0$, $^{13}$CO $J=1-0$, and $^{12}$CO $J=2-1$ transitions to look for signs of interaction between the remnant
and its neighbouring molecular environment. We also looked at archival H I, infrared, and optical data to probe different ISM
components that might also be interacting with the remnant. We conclude the following:
   \begin{enumerate}
      \item The molecular gas surrounding the western side of VRO 42.05.01 is clumpy, and rather sparse. The unusual two-shelled
      morphology of the remnant is not due to the presence of a molecular cloud in the interface between the shell and the wing.
      \item There is no physical proof, from the CO observations, that the SNR shock front has reached the observed neighbouring molecular clouds
      in the area covered by our molecular observations.
      \item There is a velocity gradient in one of the observed molecular clouds. We proposed that it was caused by a wind from VRO 42.05.01's progenitor
      star. The observed parameters correspond to a star of $12-14$~\msun.
      \item We think that the distance to the SNR was misdiagnosed in earlier papers, and propose that is $1.0\pm0.4$~kpc away, at a LSR velocity of
      $-6$~km~s$^{-1}$. This coincides with a cavity in the distribution of atomic hydrogen.
      \item There is a sheet or thick filament of dust that crosses the SNR in the interface between the shell and the wing. This feature is the most likely
      to be responsible for the two-shelled shape of the source and in particular of the sharp line between the shell and the wing.
      \item There is some infrared emission suggestive of shocked molecular clouds in regions of the remnant not covered by our IRAM observations.
      \item We speculate on the scenario that could give rise to VRO 42.05.01's shape. We favour an explanation whereby the progenitor of VRO 42.05.01 was moving
      supersonically, causing a bow
      shock and exploded on the wing side of the remnant. A reflected shock could be responsible for the triangular shape of the wing, and for the centrally-peaked
      X-ray morphology.
   \end{enumerate}

\begin{acknowledgements}
We thank the referee for his thoughtful comments which made the paper better. We also thank the IRAM
staff for their support during our observations.
The work of MA and JV is supported  by a grant from the Netherlands Research School for Astronomy (NOVA). 
This work is based on observations carried out under project number 045-18
with the IRAM 30m telescope. IRAM is supported by INSU/CNRS (France), MPG (Germany) and IGN (Spain).
The research leading to these results has received funding from the European Union's Horizon 2020 research and innovation program under RadioNet.
       This publication makes use of data products from the Wide-field Infrared Survey Explorer, 
       which is a joint project of the University of California, Los Angeles, and the Jet Propulsion Laboratory/California Institute of Technology,
      funded by the National Aeronautics and Space Administration.    
\end{acknowledgements}

\bibliography{./october}


\begin{appendix}

\section{Rms, $^{12}$CO $J=2-1$ and $^{13}$CO $J=1-0$ maps}

We present here the root mean square maps for each transition, as well as the $^{12}$CO $J=2-1$ and $^{13}$CO $J=1-0$ maps, both for the close and the distant components. 

\begin{figure*}
\centering
\includegraphics[width=0.9\textwidth]{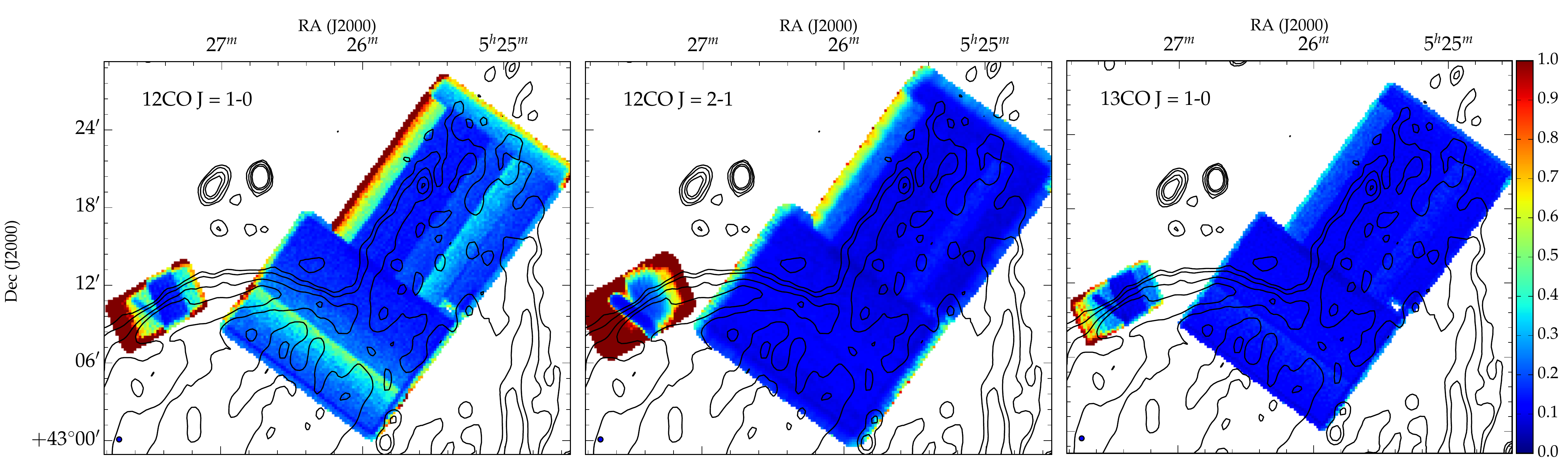}
      \caption{Root mean square noise for each of the $^{12}$CO $J=1-0$, $^{12}$CO $J=2-1$, and $^{13}$CO $J=1-0$  maps.
      All images are in the same colour scale, and the units in the colour bar are in Kelvin. Overlaid are the radio contours at 1420 MHz
      in units of brightness temperature at 4.8, 5.0, 5.5, and 6.0 K.
      The uneven rms values throughout our maps
      are due to us observing in many different weather and source elevation conditions.}
         \label{rms_noise}
\end{figure*}

   \begin{figure*}
            {\includegraphics[width=\textwidth]{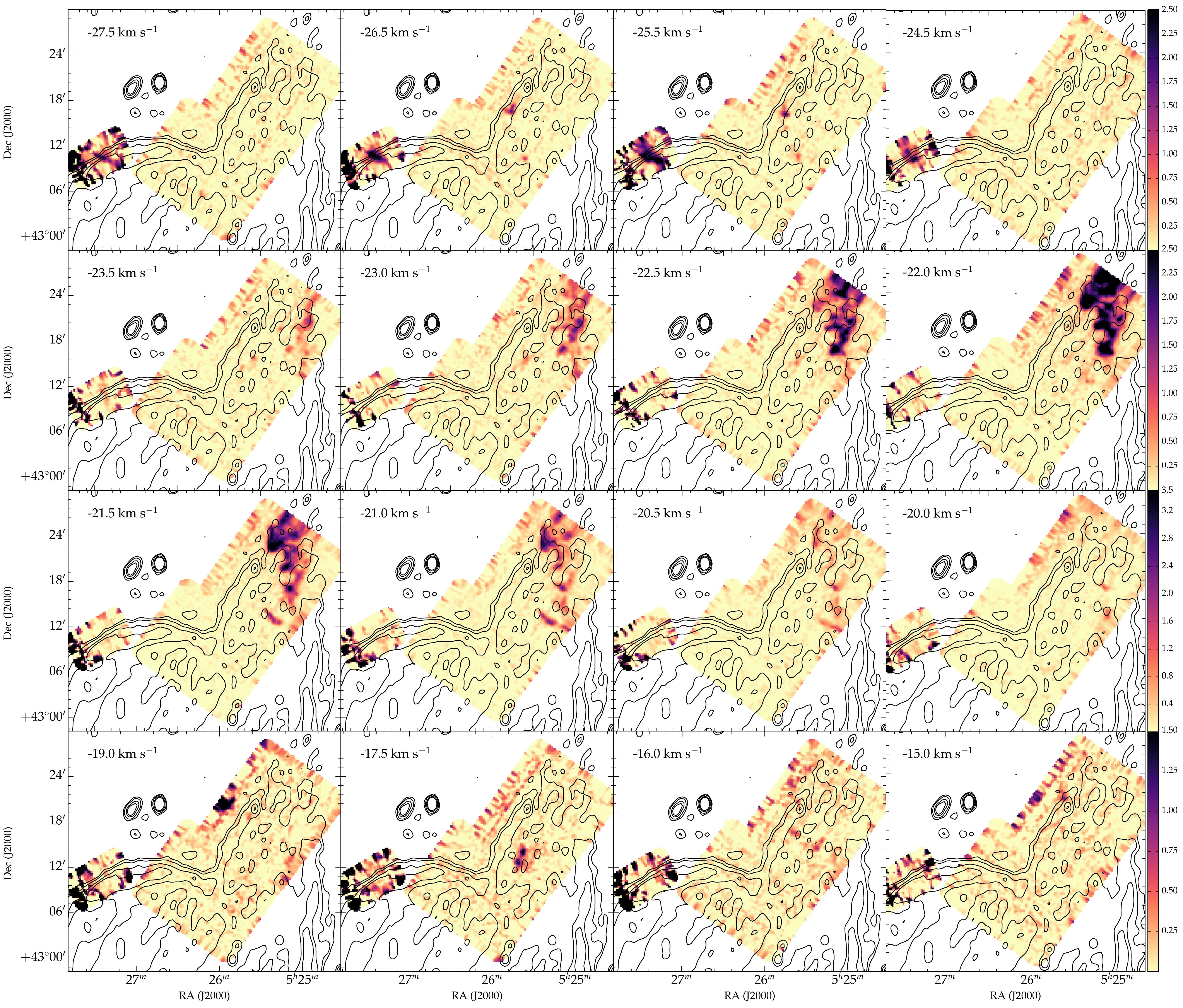}
       \caption{$T_{\mathrm{mb}}$ (K) for the $^{12}$CO $J=2-1$ line for the distant molecular cloud components, corresponding to velocities between 
      $-27.5 \, \mathrm{km\,s}^{-1}$ and $-15.0 \, \mathrm{km\,s}^{-1}$. Overlaid are the radio contours at 1420 MHz
      in units of brightness temperature at 4.8, 5.0, 5.5, and 6.0 K.
      The velocities are indicated in the upper left corner of each panel. All figures in the same row share the same colour bar, plotted at the
      left end of each row.}
         \label{12CO21}}
   \end{figure*}
   
      \begin{figure*}
            {\includegraphics[width=\textwidth]{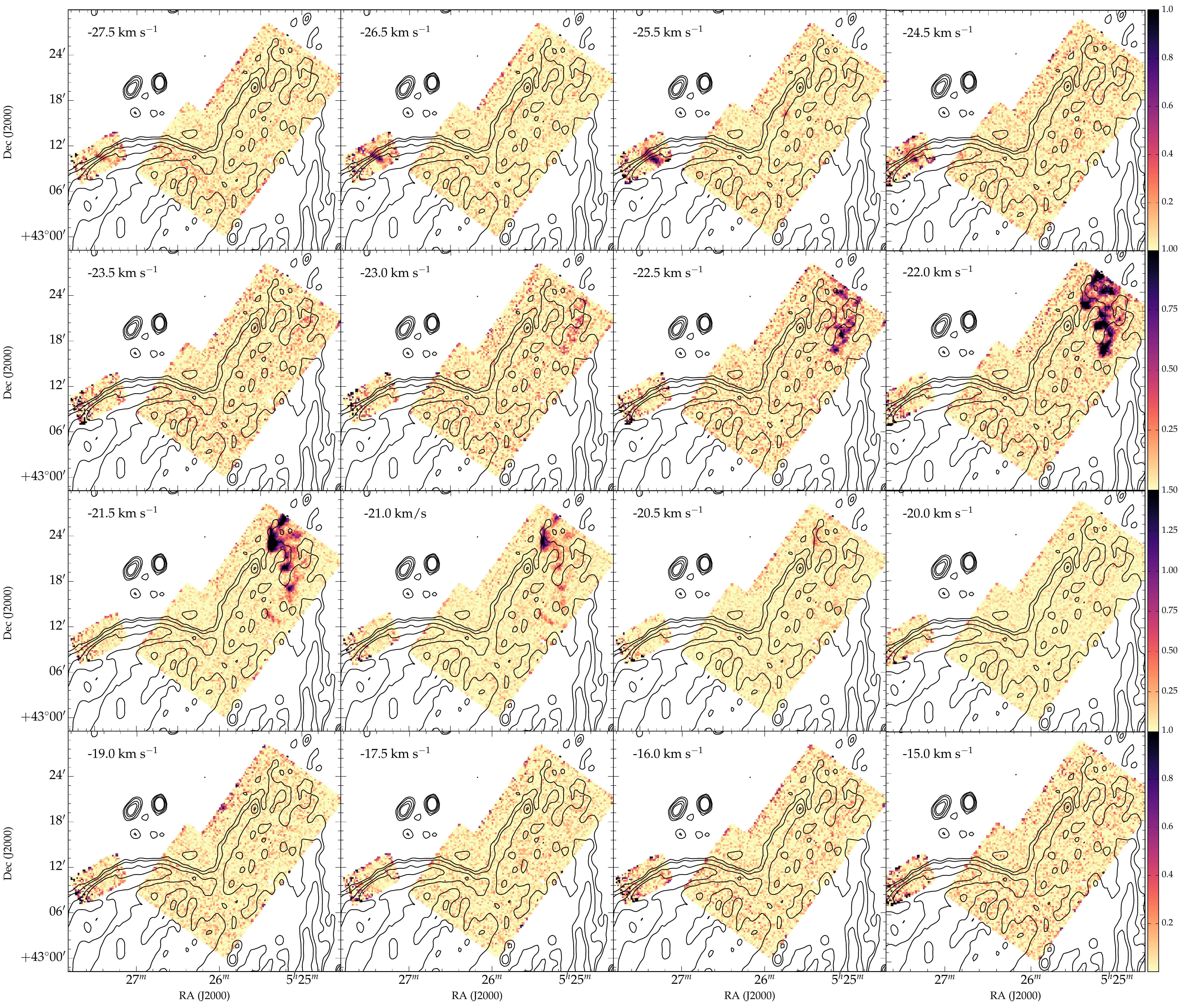}
       \caption{$T_{\mathrm{mb}}$ (K) for the $^{13}$CO $J=1-0$ line for the distant molecular cloud components, corresponding to velocities between 
      $-27.5 \, \mathrm{km\,s}^{-1}$ and $-15.0 \, \mathrm{km\,s}^{-1}$. Overlaid are the radio contours at 1420 MHz
      in units of brightness temperature at 4.8, 5.0, 5.5, and 6.0 K.
      The velocities are indicated in the upper left corner of each panel. All figures in the same row share the same colour bar, plotted at the
      left end of each row.}
         \label{13CO10}}
   \end{figure*}
   
         \begin{figure*}{
         \centering
            \includegraphics[width=0.9\textwidth]{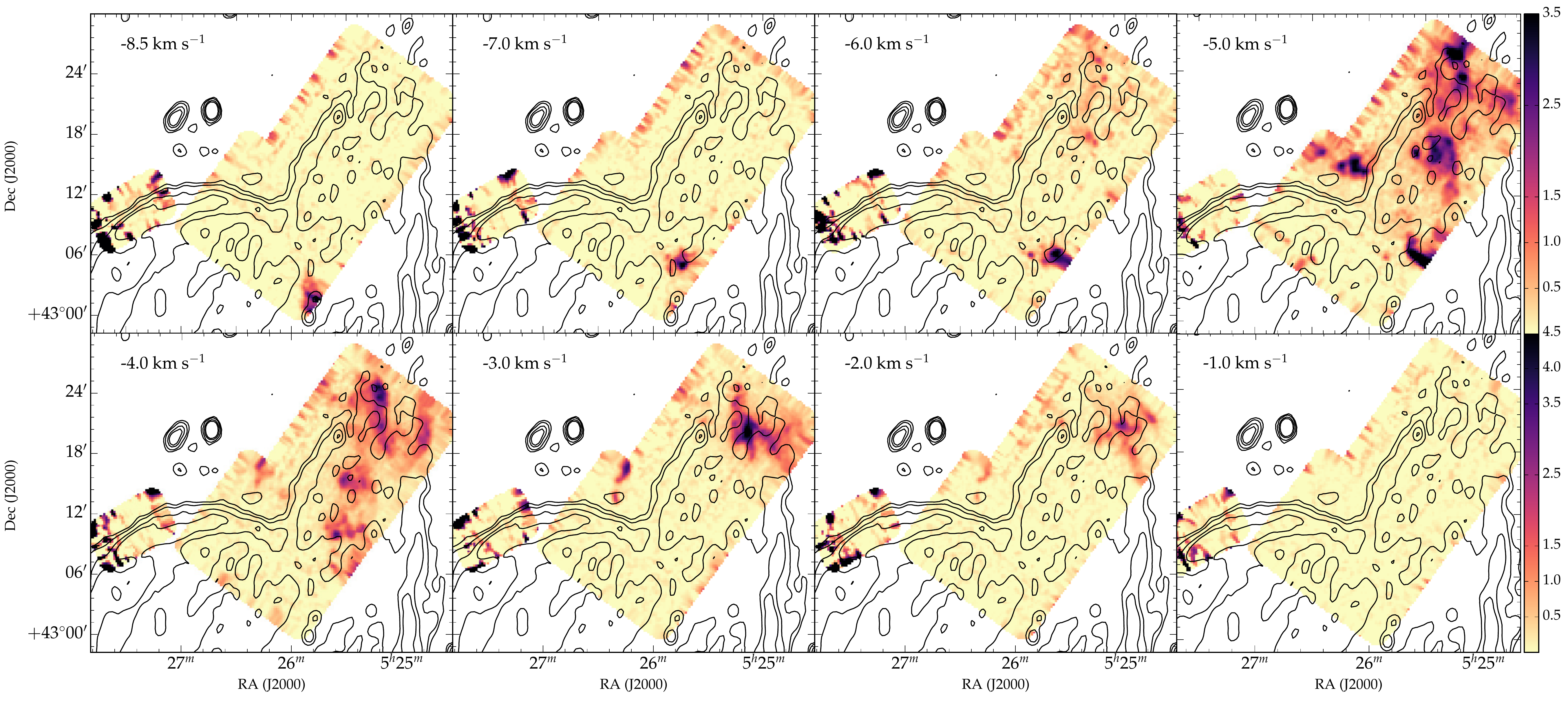}
            \includegraphics[width=0.9\textwidth]{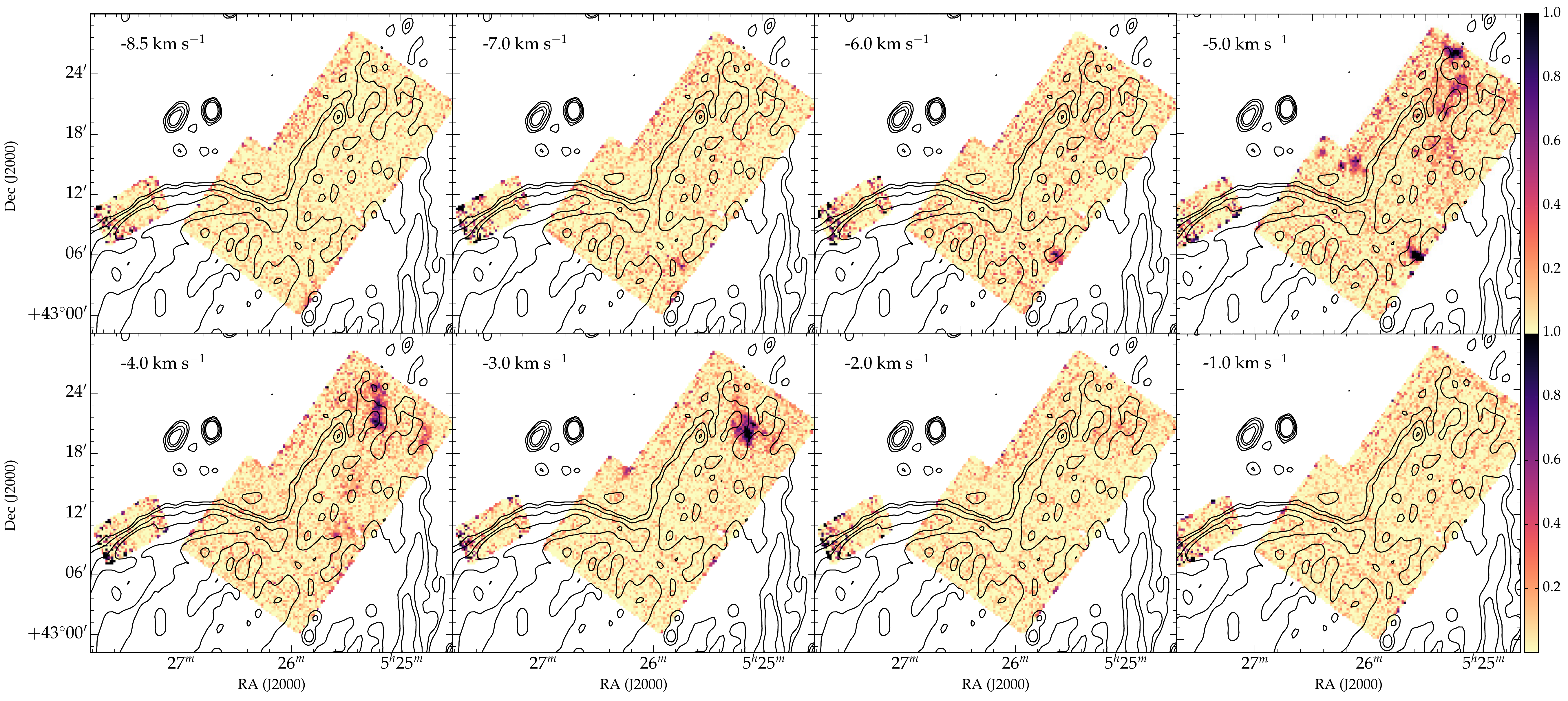}
       \caption{$T_{\mathrm{mb}}$ (K) for the  $^{12}$CO $J=2-1$ (top), and $^{13}$CO $J=1-0$ (bottom) lines
       for the close molecular cloud components, corresponding to velocities between 
      $-8.5 \, \mathrm{km\,s}^{-1}$ and $-1.0 \, \mathrm{km\,s}^{-1}$. Overlaid are the radio contours at 1420 MHz
      in units of brightness temperature at 4.8, 5.0, 5.5, and 6.0 K.
      The velocities are indicated in the upper left corner of each panel. All figures in the same row share the same colour bar, plotted at the
      left end of each row.}
         \label{local}}
   \end{figure*}

\end{appendix}

\end{document}